\documentstyle[12pt]{report}

\begin{document}

\baselineskip=18pt
SU-4240-525 \\
\begin{flushright}
November 1992\\
\end{flushright}
\vskip1.0cm
\centerline {\bf\Large EFFECTIVE HADRON DYNAMICS: FROM}
\vskip.5cm
\centerline {\bf\Large MESON MASSES TO THE PROTON SPIN PUZZLE}
\vskip.5cm
\centerline {\large J. Schechter  \small and \large A. Subbaraman}
\vskip.3cm
\centerline {\it Department of Physics, Syracuse University,}
\centerline {\it Syracuse, NY 13244-1130, USA}
\vskip.3cm
\centerline{ and}
\vskip.3cm
\centerline{\large H. Weigel}
\vskip.3cm
\centerline{\it Institut f\"ur Theoretische Physik, Universit\"at T\"ubingen,}
\centerline{\it 7400 T\"ubingen, Germany }
\vskip1.5cm
\normalsize
\centerline {\large Abstract}
\vskip.3cm
We construct a three flavor chiral Lagrangian of pseudoscalars and vectors with
special emphasis on the symmetry breaking terms.  Comparing tree level
two and three point functions with experiment allows us to first, fix the
parameters of the model (including the light quark mass ratios) and second,
to predict $m(K^{*+})-m(K^{*\circ}),\, \Gamma(K^*\rightarrow K\pi)$ and
$\Gamma(\phi\rightarrow K {\overline K})$. The last mentioned quantities
come out reasonably well, in contrast to an ``ordinary" $SU(3)$ treatment.
For this purpose we need ``second order" symmetry breakers involving the
vector fields analogous to those needed for the chiral perturbation theory
program with only pseudoscalars. An improved description of the
$\eta-\eta^\prime$ system is also given.   We then use the soliton sector
of this improved chiral Lagrangian to investigate some aspects of
baryon physics which are especially sensitive to symmetry breaking.
For this purpose a fairly elaborate ``cranking" techinque is employed
in connection with the collective Hamiltonian.  In addition to the ``strong"
baryon mass spectrum a careful investigation is made of the
non-electromagnetic part of the neutron-proton mass difference.  This work
is needed to improve our previous estimates concerning the two component
approach to the ``proton spin" puzzle.  We find that both the ``matter"
and ``glue" contributions are small but they do tend to cancel each other.

\newpage

\stepcounter{chapter}
\setcounter{equation}{0}
\leftline{\large\it 1. Introduction}
\vskip.5cm

In this paper we first discuss a three flavor chiral Lagrangian of
pseudoscalars and vectors with special emphasis on symmetry breaking. The
soliton sector is then used to calculate the non-electromagnetic part of the
neutron-proton mass difference as well as the nucleon matrix element
of the axial
singlet current. Our motivation is to update some earlier papers [1,2,3]
(which should be consulted for adequate references to background work).
However, this updating has turned out to be non-trivial and to yield some
results which may be of general interest.

The proton matrix element of the singlet axial vector current has recently
attracted a lot of attention because the EMC experiment [4] indicates that
it is close to zero at $q^2=0$. This is called the ``proton spin puzzle"
[5,6] since the matrix element should equal unity (twice the proton spin)
in the naive non-relativistic quark model. Amusingly, the simplest Skyrme
model of pseudoscalars only does predict zero [7].  One might therefore
take the point of view that the simple Skyrme model be considered a kind of
first approximation to low energy dynamics.  We have argued [1] that more
is needed because additional ``short distance" information is required to
adequately explain the neutron-proton mass difference. In fact, the
calculations of the $n-p$ mass difference and of the axial proton matrix
element
are closely linked.  Both require (at the dominant two flavor level) the
``excitation" of the $\eta$ meson field; the vanishing of one quantity
implies the vanishing of the other.  With vector mesons supplying the
short distance component it is actually possible to explain [1] the
neutron-proton mass difference and obtain [2,8] a suitably small value
for the axial singlet matrix element.  Another aspect of the proton spin
puzzle which this kind of model might illuminate is the so-called ``two
component decomposition".  This is an attempt to make the small value of
the axial singlet matrix element plausible from the QCD parton model point
of view.  The original idea [9] of decomposing the axial singlet current
into ``matter" and ``glue" parts has been criticized [5,10] as not being
gauge invariant.  However, a way to overcome this objection - by looking
instead at a two component Goldberger-Trieman type of relation - has been
suggested by Shore and Veneziano [11].  When this mechanism is implemented
[3] in the meson Lagrangian it turns out to result, at the soliton level,
in an extra contribution to the $n-p$ mass difference proportional to the
``glue" component of the axial singlet matrix element.  A numerical estimate
suggested [3] that the glue component, as well as the matter component, was
small.

Now, most of the above work consisted of {\it {calculations}} for the two
flavor part but merely {\it {estimates}} for the effects of including the
third flavor.  In this paper we discuss the calculations using the full
three flavor model throughout.  It should be remembered that the $n-p$
mass difference is driven by the symmetry breaking terms in the meson
Lagrangian so it is especially important to treat these carefully.  The
same care is warranted for treatment of symmetry breaking in the collective
Hamiltonian which describes the soliton sector of the theory.

The first step of studying symmetry breaking in the meson Lagrangian is
discussed in sections 2, 3 and 4.  We want to have sufficient symmetry
breaking terms to be able to explain at least the pseudoscalar decay
constants and masses as well as the mass spectrum of the vectors and the
$1^- \rightarrow 0^- \,0^-$ decay amplitudes.  How to do this for the
pseudoscalars is well known.  In the chiral perturbation theory (CPT) program
[12] a list of symmetry breakers depending on the mass matrix in the
underlying theory of QCD gives sufficient flexibility to fit the $0^-$ decay
constants and masses (apart from the $\eta - \eta^\prime $ system for which
a special treatment based on the $U(1)_A$ anomaly is required).  There are
also some loop diagrams resulting in ``chiral logs" but these are numerically
small for the indicated choice of scale .  On the other hand, it has been
generally felt that no such elaboration is required for understanding
symmetry breaking in the vector multiplet.  The vectors are typically
considered [13] the most ``normal" multiplet.  Here we emphasize that this is
not really the case.  Both the isospin splittings and $1^- \rightarrow 0^-
\,0^-$ decay amplitudes show large deviations from the simple $SU(3)$
predictions.  We find that these {\it {can}} be understood if suitable vector
symmetry breaking terms which are the analogs of the dominant ones for the
pseudoscalars are included.  What it boils down to is the use of the
symmetry breakers which satisfy Okubo's [14]  original form of the quark
line rule [14,15], except for the $0^-$ isosinglet channel.  An improved fit
for that channel is also discussed here in section 4.  In section 7 we make
some further remarks related to developing a larger analog of the CPT program
which would also include vectors and presumably describe low energy hadron
physics up to around 1 GeV.

Returning to the main track of this paper, we describe, in section 5, the
calculation of the $n-p$ mass difference using the improved meson Lagrangian.
A number of improvements, which have been described in detail elsewhere, are
included for treating the $SU(3)$ collective Hamiltonian of the soliton sector.
In the first place, the collective Hamiltonian is diagonalized exactly using
the Yabu-Ando technique [16].  In addition, ``cranking" corrections are
included to allow for ``centrifugal" effects with the rotating Skyrmions.
This means that while the classical soliton involves the  $ \pi^a$,
$\omega_\circ$ and  $\underline {\rho}^a$ fields, once it starts
rotating, fields
for $\eta,\eta^\prime, K, {\overline K}, \underline {\omega},
 \rho_\circ^a, K^*_\mu$
and ${\overline K}^*_\mu$ get ``excited" [17-20] with amplitudes proportional
to the rotational ``angular velocities".
This yields improved results for ``strong" baryon mass differences,
static properties, etc. We find that the model which has been thus improved
over that of ref. 1 continues to give a reasonable description of the
non-electromagnetic part of the $n-p$ mass difference.  What changes is the
percentage of this difference due to the excitation of the $\eta$ and/or
$\eta^\prime$ fields.  Previously [1] it was around $70\%$ but is now reduced
to about $20\%$ for reasons discussed in section 5.  The application of this
calculation to determining the ``glue" part of the proton's axial singlet
matrix element is given in section 6.  We continue to find that this
contribution is small.  However,  because there is a smaller $\eta$
contribution to the $n-p$ mass difference, the accuracy of this statement
is actually decreased somewhat.

\vskip.5cm

\stepcounter{chapter}
\setcounter{table}{0}
\setcounter{equation}{0}
\leftline{\large\it 2. Terms of the Lagrangian}
\vskip.5cm

Here we collect together, for convenience, the  {\bf a.} chiral invariant,
 {\bf b.} flavor symmetry breaking but quark line rule conserving and
{\bf c.} quark line rule violating terms of the pseudoscalar$-$vector
effective Lagrangian.  Some discussion will be given of why we have chosen
to include the terms listed, but not others.

The dynamical degrees of freedom are the elements of the $3\times 3$ matrix of
the pseudoscalar nonet, $\phi$ and the $3\times 3$ matrix of the vector nonet
$\rho_{\mu}$. We need the unitary matrices
\begin{eqnarray}
U = e^{2i \phi /F_\pi}, \quad \xi = U^{1/2} = e^{i \phi /F_\pi},
\end{eqnarray}
where $F_{\pi}$ is a {\it {bare}} pion decay constant. The vector mesons,
which also transform non-linearly under chiral $U(3) \times U(3)$, are related
to auxiliary linearly transforming ``gauge fields" $A_{\mu}^L$ and $A_{\mu}^R$
by [21]
\begin{eqnarray}
A_{\mu}^L = \xi \rho_{\mu} \xi^{\dagger} + {i\over \tilde g}\xi
\partial_{\mu} \xi^{\dagger},  \ \
A_{\mu}^R = \xi^{\dagger} \rho_{\mu} \xi + {i\over \tilde g}\xi^{\dagger}
\partial_{\mu} \xi ,
\end{eqnarray}
where $\tilde g$ is a {\it {bare}} $\rho \phi \phi$ coupling constant.

Note that we have included only the lowest lying s-wave $q\overline q$
bound states of QCD as our dynamical fields.

\vskip.5cm
\leftline {{\bf a.} {\it {Chiral invariant terms}}:[21-23]}
\vskip.3cm
These include the kinetic piece for the vectors:
\begin{eqnarray}
-{1 \over 4} Tr[F_{\mu \nu}(\rho)F_{\mu \nu}(\rho)] , \quad
F_{\mu \nu}(\rho) = \partial_{\mu}\rho_{\nu} - \partial_{\nu}\rho_{\mu}
- i{\tilde g} [\rho_{\mu},\rho_{\nu}] ,
\end{eqnarray}
and pseudoscalar kinetic plus vector interaction pieces:
\begin{eqnarray}
\lefteqn{- {m_v^2(1+k) \over 8k} Tr(A_\mu^L A_\mu^L
+ A_\mu^R A_\mu^R) +
  {m_v^2(1-k) \over 4k}Tr(A_\mu^L  U A_\mu^R U^\dagger)} \nonumber \\
 =& -&{1\over2}m_v^2 Tr(\rho_\mu \rho_\mu)-{im_v^2 \over 2{\tilde g}}
 Tr[\rho_\mu(\partial_\mu \xi \xi^\dagger + \partial_\mu \xi^\dagger \xi)]
 -{F_\pi^2 \over 4}(1+k) Tr (\partial_\mu \xi \partial_\mu \xi^\dagger)
 \nonumber \\
 &-&{F_\pi^2 \over 4}(1-k) Tr(\xi^\dagger \partial_\mu \xi^\dagger
\xi \partial _\mu \xi) ,
\end{eqnarray}
where $m_v$ is a {\it{bare}} vector meson mass and $k$ is a dimensionless
constant chosen as $k=(m_v/F_\pi \tilde g)^2$ to insure correct normalization
of the pseudoscalar kinetic term. Please notice that we are using the
$``x_4 = it"$ Euclidean-type metric convention.

There are also important interaction terms proportional to the Levi-Civita
symbol, $\epsilon_{\mu \nu \alpha \beta}$. To write these compactly it is
convenient to use a form notation with one-forms $\alpha = dUU^{-1}$ and
$A^L$ and to write the action terms (rather than the Lagrangian density terms):
\begin{eqnarray}
\Gamma_{WZ}(U) + \int Tr&[ic_1 (A^L\alpha^3) + c_2 (dA^L\alpha A^L
 - A^L\alpha dA^L + A^L\alpha A^L \alpha) \nonumber \\
 \phantom{aaaaaaaaaaaaaaaaa} +& c_3(-2i(A^{L})^3\alpha +
 {1 \over \tilde g} A^L \alpha A^L \alpha)]\phantom{aaaaaaaaaaa} ,
\end{eqnarray}
where $c_1, c_2, c_3$ are constants discussed in [20-23] while $\Gamma_{WZ}(U)$
is the Wess-Zumino-Witten term [24]:
\begin{eqnarray}
\Gamma_{WZ}(U) = {-i \over {80\pi^2}} \int\limits_{{\cal M}^5} Tr(\alpha^5) .
\nonumber
\end{eqnarray}

Eqns (2.3), (2.4) and (2.5) comprise a minimal (though nevertheless complicated
in detail) model for pseudoscalars and vectors [25]. One might imagine
extending
it by systematically adding terms with two more derivatives.  This would be
analogous to the usual chiral perturbation theory (CPT) program [12] with only
pseudoscalars included.  It should be remarked, however, that a good deal of
the
four derivative structure of the pseudoscalar-only effective Lagrangian can be
obtained by ``integrating out" the vectors of a {\it {minimal}}-type
pseudoscalar-vector effective Lagrangian.

\vskip .5cm
\leftline {{\bf b}.  {\it {Symmetry breaking terms}}:}
\vskip.3cm
In this ``updating" of section II of ref. [1]  we will include only quark-line
rule conserving terms, $i.e.$ those which can be written as a {\it {single}}
trace in flavor space.

The current quark mass terms in the fundamental QCD Lagrangian may be written
as
\begin{eqnarray}
{\cal L}_{mass} = - {\hat m} \overline q {\cal M}q ,
\end{eqnarray}
where $q$ is the column vector of up, down and strange quark fields,
$\hat m \equiv (m_u + m_d)/2$ and ${\cal M}$ is a dimensionless, diagonal
matrix which can be expanded as follows:
\begin{eqnarray}
{\cal M} = y\lambda_3 + T + xS,
\end{eqnarray}
with $\lambda_3 = diag(1,-1,0),\, T=diag(1,1,0)$ and $S=diag(0,0,1)$. $x$ and
$y$ are the quark mass ratios
\begin{eqnarray}
x = {m_s \over \hat m}, \quad  y = -{1\over2} \left({{m_d - m_u} \over \hat m}
\right) .
\end{eqnarray}
It is also convenient to define [13]
\begin{eqnarray}
R = {{m_s - \hat m} \over {m_d - m_u}} = {{1-x} \over 2y} .
\end{eqnarray}
The significance of $R$ is that it can, in principle, also be determined from
the rather well-known masses of the ${1\over2}^+$ baryon octet.  We have
introduced the redundant notations above because of the importance of
${\cal M}$
in (2.7), to which all the symmetry breakers are taken to be proportional.  We
then write for the quark-line-rule conserving but flavor symmetry breaking
terms in the effective Lagrangian:
\begin{eqnarray}
&&Tr \big( {\cal {M}} \big[\alpha^{\prime}(A_\mu^L U A_\mu^R +
A_\mu^R U^{\dagger}
 A_\mu^L) + \beta^{\prime}(\partial_{\mu}U\partial_{\mu}U^{\dagger}U +
 U^{\dagger}\partial_\mu U\partial_{\mu}U^{\dagger}) \nonumber \\
&&\qquad+\gamma^{\prime}(F_{\mu\nu}^LUF_{\mu\nu}^R +
  F_{\mu\nu}^{R} U^{\dagger}F_{\mu\nu}^{L})
 +  \delta^{\prime}(U+U^{\dagger}-2) \big]  \nonumber \\
&&\qquad+\lambda^{\prime 2} \big[ {\cal M}U^{\dagger}{\cal M} U^{\dagger}+{\cal
 M}U{\cal M}U-2{\cal M}^2 \big] \big) , 
\end{eqnarray}
where $F_{\mu\nu}^{L,R} = \partial_{\mu}A_{\nu}^{L,R}-i{\tilde g}
  [A_{\mu}^{L,R},A_{\nu}^{L,R}]$ and
$\alpha^{\prime},\beta^{\prime},\gamma^{\prime},
\delta^{\prime},\lambda^{\prime}$ are the symmetry breaking coefficients.
The parameters $\alpha^{\prime \prime}, \beta^{\prime \prime}$ etc.
introduced in ref. [1] may be related easily to $\alpha^{\prime},
\beta^{\prime}$ etc. via:
\begin{eqnarray}
x = \alpha^{\prime \prime} / \alpha^{\prime}  =
    \beta^{\prime \prime} / \beta^{\prime}  = etc., \quad
y = \alpha / \alpha^{\prime} =  \beta / \beta^{\prime} = etc.
\end{eqnarray}

Our motivations for including the terms shown in (2.10) are as follows. First,
the $\delta^{\prime}$ term is the standard one which splits the pseudoscalar
masses.  Next, the $\alpha^{\prime}$ term is the one with the correct chiral
properties which splits the vector masses.  However, as pointed out in ref.
[1],
the $\alpha^{\prime}$ term also gives objectionably large derivative-type
symmetry breaking for the pseudoscalars.  This is partially canceled by the
pure pseudoscalar derivative symmetry breaker given by the $\beta^{\prime}$
term.  The $\gamma^{\prime}$  term yields a derivative-type symmetry breaker
for the vectors too.  Note that the $\beta^{\prime}$  term is the \#5 type in
the CPT classification [12]. The $\lambda^{\prime2}$ term, which involves two
powers of ${\cal M}$, is the \#8 type.  It is argued to be the same order as
the
$\beta^{\prime}$  term.  Furthermore, it is the only remaining manifestly
quark-line-rule conserving symmetry breaker to ``second order".  From a
practical point of view the $\lambda^{\prime2}$ term also enables us, as
we shall see, to vary $x$ in (2.8) while keeping the pseudoscalar masses and
decay constants fixed at their measured values.  In the CPT program, the
determination of $\lambda^{\prime2}$ is subject to the Kaplan-Manohar
ambiguity [26]; we shall discuss this point later.

\vskip.5cm
\leftline {{\bf c.} {\it {Quark-line-rule violating and symmetry breaking
  terms}}:}
\vskip.3cm
One of the remarkable features of low energy dynamics is the success of the
OZI or quark-line rule [14,15].  There is however a conspicuous
exception associated
with the interactions of the pseudoscalar singlet particles ($\eta,
\eta^{\prime}...$); the physics of this is denoted  as the ``$U(1)$ problem".
In a rough way, this picture is reflected in the CPT fit of ref. [12]. There
the
OZI rule violating terms of  \#4 type ($\sim Tr(\partial_{\mu}U
\partial_{\mu}U^{\dagger})\,Tr({\cal M}(U+U^{\dagger}))$ ) and of  \#6 type
($\sim [Tr({\cal M}(U+U^{\dagger}))]^2$ ) are both claimed to be negligible.
On the other hand the \#7  type OZI rule violating symmetry breaker
($\sim [Tr({\cal M}(U-U^{\dagger}))]^2$ ) is claimed to be non-negligible.
We shall see that the latter is naturally associated with
pseudoscalar singlet particle
interactions.  Incidentally note that we have now accounted for all the
symmetry breaking terms involving only pseudoscalars which appear in the second
order CPT list.  To these we have added the $\alpha^{\prime}$ and
$\gamma^{\prime}$  terms in (2.10), which involve the vectors.

Here, we shall write down essentially just those OZI rule violating terms
needed to fit the $\eta$ and $\eta^{\prime}$ masses and two photon decay
widths in an effective chiral Lagrangian framework.  In order that the $U(1)$
anomaly equation for the axial singlet current, $\partial_{\mu}J_{\mu}^5=G$,
with $G=\partial_{\mu}K_{\mu}, \,K_{\mu}=
{{-3ig_{QCD}^2} \over{4\pi^2}}\epsilon_
{\mu\beta\gamma\delta}Tr(A_\beta \partial_\gamma A_\delta
- {2ig_{QCD} \over 3} A_\beta A_\gamma A_\delta)$
in terms of the QCD gauge fields, be obeyed
in the massless limit it is convenient to introduce $G$ as an ``auxiliary"
field in the effective Lagrangian.  No kinetic term is to be written for
$G$ so it gets eliminated by its equation of motion.   This procedure might
be considered as an effective way of integrating over instanton field
configurations.  We thus add the following terms to our effective Lagrangian:
\begin{eqnarray}
{1\over \kappa}G^2 &+& {i\over{12}}G\,\ln (\det U/ \det U^\dagger)
+n\,Tr(\alpha_\mu)Tr(\alpha_\mu)
+ iG\,\epsilon^\prime\, Tr[{\cal M}(U- U^\dagger)] \nonumber \\
&+&iG\,\zeta^\prime\, Tr[{\cal M}(A_{\mu}^L U A_{\mu}^R - A_{\mu}^R
 U^\dagger A_{\mu}^L)],
\end{eqnarray}
where $\kappa, n, \epsilon^\prime$ and $\zeta^\prime$ are parameters.  The
first two terms in (2.12) are the standard ones for mocking up the $U(1)$
anomaly and giving mass to the $\eta^\prime$ meson [27].  The third term, which
gives us the freedom to adjust the $\eta^\prime$ ``decay constant", and the
fourth term which is a symmetry breaker of ``instanton-induced" type, have
been mentioned before [28].  The fifth term is an ``instanton-induced" symmetry
breaker involving vectors; since it leads only to four and higher point
interactions we shall neglect it for our present purposes by setting
$\zeta^\prime = 0$.  Now $G$ appears only in (2.12).  It therefore satisfies
the equation of motion
\begin{eqnarray}
G = {\kappa \over4}\chi - {{i\kappa \, \epsilon^\prime} \over2}
 Tr[{\cal M}(U-U^{\dagger})],
\end{eqnarray}
where the mathematical ( $SU(3)$ singlet) $\eta^\prime$ field is defined by
$\eta^\prime={{\sqrt{3}}\over2}F_\pi \chi$,  $\chi$ being gotten from the
decomposition $U=e^{i\chi}\tilde U$ with $\det \tilde U =1$.  $G$ in (2.13)
should be substituted into (2.12).  Among other things it gives a term of
\#7 type in the CPT list.  We have no special reason, on the other hand, to
include a term of  \#6 type.

An additional term involving $G$ will be considered later in connection with an
application to the ``proton spin" puzzle.

\vskip.5cm
\leftline{ {\bf d.} {\it {Remarks}}:}
\vskip.3cm
The total effective action of pseudoscalars and vectors is taken to be
\begin{eqnarray}
\Gamma_{eff} = \int [\,(2.3) + (2.4) + (2.10) + (2.12)]d^4x + (2.5).
\end{eqnarray}
Symmetry breaking terms proportional to $\epsilon_{\mu\nu\alpha\beta}$ are
being neglected here for simplicity (they do not contribute to most of the
processes which will be discussed in the present paper).  Similarly, OZI
rule violation for particles other than the pseudoscalars will mostly be
neglected.
\vskip.5cm

\setcounter{equation}{0}
\setcounter{table}{0}
\stepcounter{chapter}
\leftline{\large\it 3. Physical quantities at the OZI rule conserving level}
\vskip.5cm
We will find the parameters of the effective Lagrangian by comparing the
tree level 2 and 3 point functions with experiment.  There is a natural
separation into the properties of the pseudoscalar singlets (which we discuss
in the next section) and the other particles whose properties follow just
from the OZI rule conserving terms.

Compared to the discussion in Sec. II of ref. [1], there is the difference
that the $\lambda^\prime$ term is now included and also the
$\gamma^\prime$ term is not set to zero.  The latter results in a very
substantial improvement in the predictions for the properties of the
$K^\star$ mesons and gives a different fit for the fundamental mass
ratios $x$ and $y$ of the light quarks.

First, expanding out the kinetic terms shows we should renormalize the fields
as (taking typical examples):
\begin{eqnarray}
\pi^+ = Z_{\pi}\phi_{12} ,\, &  K^+ = Z_K \phi_{13},\, &  \rho^{+}_{\mu}=
Z_\rho \, \rho_{12\mu}, \nonumber \\
K^{\star +}_{\mu} = Z_{K^\star}\, \rho_{13\mu},\,  &
\omega_\mu= Z_\omega(\rho_{11\mu}+\rho_{22\mu})/\sqrt{2}, \, &
\phi_\mu = Z_\phi \, \rho_{33\mu},
\end{eqnarray}
where
\begin{eqnarray}
Z_\pi= \big[1+{4\over{F_{\pi}^2}}({{\alpha^{\prime}}\over{{\tilde g}^2}}-4\beta
^\prime) \big]^{1/2}, \>  &
Z_K=\big[1+{2\over{F_{\pi}^2}}(1+x)({{\alpha^\prime}\over{{\tilde g}^2}}-
4\beta^\prime) \big]^{1/2} , \nonumber \\
Z_\rho=Z_\omega=(1-8\gamma^\prime)^{1/2}\phantom{aa}\quad, &
Z_K^\star = \big[1-4\gamma^\prime(1+x) \big]^{1/2},\qquad \nonumber \\
Z_\phi = (1-8x\gamma^\prime)^{1/2}.\phantom{aaaaaaaa} & \phantom{AAAAA}
\end{eqnarray}

Next, the computation of the ``Noether" currents of this model leads to the
identification of the physical pion and kaon decay constants $F_{\pi p}$
and $F_{Kp}$ as:
\begin{eqnarray}\label{somename}
F_{\pi p}=Z_\pi F_\pi, \quad F_{Kp}=Z_K F_\pi \>.
\end{eqnarray}

Expanding out the pseudoscalar mass terms yields, in the isospin limit,
\begin{eqnarray}
m_{\pi}^2 = {8\over{F_{\pi p}^2}}(\delta^\prime+4\lambda^{\prime2}), \quad
m_{K}^2= {4\over{F_{K p}^2}}[\delta^\prime(1+x)+2 \lambda^{\prime2}
(1+x)^2].
\end{eqnarray}
Note that, as pointed out in Sec.2, if $\lambda^{\prime2}$ were not present
the two equations in (3.4) would lead to a fixed value for $x=m_s/{\hat m}$.
However it turns out to be very useful to explore the physical situation for
various values of $x$.  The vector meson masses in the isospin limit are:
\begin{eqnarray}
m_{\rho}^2=m_{\omega}^2=(m_v^2-2\alpha^\prime)/Z_{\rho}^2, &
m_{K^\star}^2=[m_v^2-2\alpha^\prime(1+x)]/Z_{K^\star}^2, \nonumber \\
m_{\phi}^2= (m_v^2-4x\alpha^\prime)/Z_{\phi}^2 \phantom{aaa} & \phantom{a}.
\end{eqnarray}

Next, let us turn to the mass splittings between members of the same
iso-multiplet.  We are not computing the photon-exchange contributions here
so the prediction will be for the {\it{non-electromagnetic}}
part of the iso-splittings.  We then find {\footnote{Only 3.8 was used for
fitting in ref. [1].  The extra factor $Z_{\phi}^{-2}$ makes negligible
difference. Note that the $\lambda^{\prime2} \rightarrow 0$ limit of (3.6)
quoted, but not used for fitting, should be replaced by the present formula.}}
\begin{eqnarray}
[m(K^\circ)-m(K^+)]_{non-EM} &=& {y\over{F_{Kp}^2m_K}}[2m_K^2({{\alpha^\prime}
\over{{\tilde g}^2}}-4\beta^\prime) -4\delta^\prime-16\lambda^{\prime2}(1+x)],
\end{eqnarray}
\begin{eqnarray}
[m(K^{\circ *})-m(K^{+ *})]_{non-EM} &=& y(-4\gamma^\prime m_{K^*}^2
+2\alpha^\prime)/ m_{K^*}Z_{K^*}^2 ,
\end{eqnarray}
\begin{eqnarray}
[M_{\rho\omega}]_{non-EM} &=& -y(2\alpha^\prime -4\gamma^\prime m_{\rho}^2)/
m_\rho Z_{\rho}^2   .
\end{eqnarray}
(The $\rho^\circ-\omega$ transition mass, $M_{\rho\omega}$ is defined in terms
of the effective term in the Lagrangian:
$-2m_{\rho}M_{\rho\omega}\rho_{\mu}^{\circ}\omega_\mu$ ).

Finally, consider the $\rho\phi\phi$ coupling constants.  The $\rho\pi\pi$
coupling constant, $g_{\rho\pi\pi}$ may be defined in terms of the
$\rho \rightarrow 2\pi$ width, $\Gamma(\rho \rightarrow 2\pi)$ as
\begin{eqnarray}
\Gamma(\rho \rightarrow 2\pi)={{g_{\rho\pi\pi}^2\,|{\underline q}(\rho)|^3}
\over{12\pi m_{\rho}^2}} ,
\end{eqnarray}
where ${\underline q}(\rho)$ is the momentum of the final pion in the $\rho$
rest frame.   In our model,
\begin{eqnarray}
g_{\rho\pi\pi}={{m_v^2+4\alpha^\prime}\over{{\tilde g}F_{\pi p}^2
Z_\rho}}  .
\end{eqnarray}
For the other vector decays into two pseudoscalars we find
\begin{eqnarray}
{{\Gamma(K^\star \rightarrow K\pi)}\over{\Gamma(\rho \rightarrow 2\pi)}} &=&
{3\over4}\left({m_\rho\over m_{K^\star}}\,{F_{\pi p}\over F_{Kp}}\,{Z_\rho\over
Z_{K^\star}}\right)^2 {{|{\underline q}(K^\star)|^3}\over
{|{\underline q}(\rho)|^3}} , \nonumber \\
{{\Gamma(\phi \rightarrow K{\overline K})}\over{\Gamma(\rho \rightarrow 2\pi)}}
 &=& \left({m_\rho\over m_{\phi}}\,\left({F_{\pi p}\over F_{Kp}}\right)^2\,
{Z_\rho\over
Z_{\phi}}\right)^2{{|{\underline q}(\phi)|^3}\over{|{\underline q}(\rho)|^3}},
\end{eqnarray}
where ${\underline q}(K^\star)$, for example,  denotes the {\it {
pseudoscalar}} momentum in the $K^\star$ rest frame.  Also, the small OZI
rule violation for $\phi$ decays was neglected in the second part of (3.11),
as well as in the mass formula (3.5).

Now let us consider the experimental determination of the various quantities
introduced.  This is actually the heart of the matter and contains several
very interesting qualitative features.  In the first place this determination
yields estimates of the fundamental quark mass ratios $x=m_s/{\hat m}$ and
$y=(m_u-m_d)/(m_u+m_d)$ in our framework.  The dependent quantity
$R=(1-x)/2y$ is also relevant since it can be separately determined from
consideration of other particle multiplets (like the baryon octet).  For
comparison we mention the results obtained in earlier work in this model [1]
in which $\gamma^\prime=\lambda^{\prime2}=0$ and wherein $M_{\rho\omega}$
was the only experimental isospin violating quantity used for the fit:
\begin{eqnarray}
x=37,\quad y=-0.36, \quad R=50 \>.
\end{eqnarray}
Let us also compare with the determination of Gasser and Leutwyler[13]:
\begin{eqnarray}
x=25.0\pm2.5,\quad y=-0.28\pm0.03, \quad R=43.5\pm2.2 \>.
\end{eqnarray}
A recent alternate approach to the CPT program by Donoghue and Wyler [29]
predicts instead
\begin{eqnarray}
 y=-0.54\pm0.09 \>.
\end{eqnarray}
Those authors do not have a similarly precise estimate of $x$ but consider
$R=32\pm5$ and hence $x \approx 36$ to be more acceptable.  In the present
paper we find a ``best fit" for
\begin{eqnarray}
x=31.5,\quad y=-0.42, \quad R=36 \>.
\end{eqnarray}
This is closer to (3.14) than to (3.13).

The main qualitative difference between our result (3.15) and (3.13) can be
roughly understood in the following way.  Gasser and Leutwyler consider the
vector meson nonet to be the most ``normal" one in the sense that old
fashioned $SU(3)$ relations can be best trusted.  Their value of $R$, for
example, is essentially obtained from the vectors.  However we find very
important $SU(3)$ violations for the vector nonet system.  They did note
that the non-electromagnetic part of  $[m(K^{\circ *})- m(K^{+*})]$, which
is related to $M_{\rho\omega}$ (their basic isospin violation input) by
$SU(3)$, could not be adequately explained but attributed this to
uncertainties in interpreting the experimental results.  Here, we find that
the experimental value for this mass splitting  can be fairly reasonably
explained if one allows non-trivial wave function renormalization  for the
$K^*_\mu$ and $\phi_\mu$ fields by keeping $\gamma^\prime \not= 0$.  This
also allows us to dramatically improve the predictions for the widths
$\Gamma(K^*)/\Gamma(\rho)$ and $\Gamma(\phi)/\Gamma(\rho)$ in (3.11).  It
is amusing to note that when $SU(3)$ was first proposed, the symmetry
prediction for $\Gamma(K^*)/\Gamma(\rho)$ worked very well.  But since then,
the measured width of the $\rho$ has increased from about 100 MeV to about
150 MeV.  The relatively large wave function renormalization for the
$K^*$ tends to restore the agreement between theory and experiment.

It is very easy to fit the symmetry breaking parameters $x,y,\alpha^\prime,
\beta^\prime,\gamma^\prime,\delta^\prime,\lambda^{\prime2}$ as well as the
``symmetric" parameters ${\tilde g}$ and $m_v^2$ to experiment
{\footnote {Input parameters: $F_{\pi p}=0.132\, {\rm {GeV}},
F_{Kp}=0.161\, {\rm {GeV}}, m_\pi=0.137\,{\rm {GeV}},m_{K^*}=0.892\,
{\rm {GeV}},m_\rho=0.768\,{\rm {GeV}},m_K=0.497\,{\rm {GeV}},
(m_{K^\circ}-m_{K^+})_{non-EM}=5.28\,{\rm {MeV}},M_{\rho\omega}=-2.65\,
{\rm {MeV}}, \Gamma(\rho \rightarrow 2\pi)=0.1491\,{\rm {GeV}},
\Gamma(K^* \rightarrow K\pi)=0.0498\,{\rm {GeV}},
\Gamma(\phi \rightarrow K{\overline K})=0.0037\,{\rm {GeV}}$.}}
if one
temporarily holds $x$ fixed and computes everything else.  First, feeding
the known values for $F_{\pi p},F_{Kp},m_\pi$ and $m_K$ into (3.4) gives
$\delta^\prime$ and $\lambda^{\prime2}$.  From $F_{Kp}/F_{\pi p}=Z_k/Z_\pi$
we next find the quantity $({\alpha^\prime \over{{\tilde g}^2}}-4\beta^\prime)$
 from (3.2).
Equation (3.6) then yields $y$ while the formulas for $m_\rho$ and
$m_{k^*}$ in (3.5) together with (3.8) give us $\alpha^\prime,\gamma^\prime$
and $m_v$.  Next the $\rho$ width via (3.9) and (3.10) gives $\tilde g$ and
finally $\beta^\prime$ is found from $({\alpha^\prime \over{{\tilde g}^2}}
-4\beta^\prime)$.
The fitted parameters as functions of $x$ are displayed in Table 3.1.
\begin{table}
\centerline{
\begin{tabular}{|c|c|c|c|c|c|c|c|c|}
\hline
$x$ & $y$ & $R$ & $\alpha^\prime({\rm {GeV}}^2)$ &
$\beta^\prime({\rm {GeV}}^2)$ & $\gamma^\prime$ & $\delta^\prime({\rm
{GeV}}^4)$
& $|\lambda^\prime|({\rm {GeV}}^2)$ & ${\tilde g}$ \\
\hline
15 & $-$0.13 & 53.5 & $-9.08\times 10^{-3}$ & $-2.59\times 10^{-4}$ &
$-1.05\times 10^{-3}$ & $3.24\times 10^{-5}$ & $1.45\times 10^{-3}$ & 3.53 \\
20 & $-0.19$ & 48.9 & $-4.71\times 10^{-3}$ & $-1.42\times 10^{-4}$ &
$4.32\times 10^{-4}$ & $3.72\times 10^{-5}$ & $9.64\times 10^{-4}$ & 3.72 \\
25 & $-0.27$ & 43.8 & $-2.03\times 10^{-3}$ & $-7.91\times 10^{-5}$ &
$1.39\times 10^{-3}$ & $3.92\times 10^{-5}$ & $6.56\times 10^{-4}$ & 3.83 \\
28 & $-0.33$ & 40.5 & $-8.72\times 10^{-4}$ & $-5.36\times 10^{-5}$ &
$1.81\times 10^{-3}$ & $3.98\times 10^{-5}$ & $5.15\times 10^{-4}$ & 3.88 \\
30 & $-0.38$ & 38.3 & $-2.25\times 10^{-4}$ & $-4.05\times 10^{-5}$ &
$2.05\times 10^{-3}$ & $4.01\times 10^{-5}$ & $4.31\times 10^{-4}$ & 3.91 \\
32 & $-0.43$ & 36.1 & $3.43\times 10^{-4}$ & $-2.89\times 10^{-5}$ &
$2.26\times 10^{-3}$ & $4.04\times 10^{-5}$ & $3.51\times 10^{-4}$ & 3.93 \\
34 & $-0.49$ & 33.9 & $8.46\times 10^{-4}$ & $-1.88\times 10^{-5}$ &
$2.45\times 10^{-3}$ & $4.06\times 10^{-5}$ & $2.71\times 10^{-4}$ & 3.95 \\
36 & $-0.55$ & 31.6 & $1.29\times 10^{-3}$ & $-9.98\times 10^{-6}$ &
$2.62\times 10^{-3}$ & $4.07\times 10^{-5}$ & $1.84\times 10^{-4}$ & 3.97 \\
38 & $-0.63$ & 29.3 & $1.69\times 10^{-3}$ & $-2.19\times 10^{-6}$ &
$2.77\times 10^{-3}$ & $4.09\times 10^{-5}$ & $4.71\times 10^{-4}$ & 3.99 \\
\hline
\end{tabular}}
\caption{ Parameters as functions of $x$ }
\end{table}

 We note
that the strength of the standard non-derivative pseudoscalar symmetry
breaker, $\delta^\prime$, does not change much with $x$.  The most dramatic
effect is the increasing importance of $\gamma^\prime$ which, from (3.2),
is seen to lead to an important wave function renormalization for the $K^*$
and $\phi$ vector masses.  When $x=25$, the values of $x$ and $y$ agree with
the Gasser-Leutwyler values in (3.13).  As $x$ increases it is seen that the
strengths $|\beta^\prime|$ and $|\lambda^\prime|$ decrease substantially.
For our  fit in (3.15), $\alpha^\prime$ has also decreased in
magnitude and has reversed sign.  Clearly $\delta^\prime$ and $\gamma^\prime$
are the dominant symmetry breaking parameters for our fit.  The value of
the ``gauge" coupling constant $\tilde g$ does not change much with $x$.

We now have four predictions which are given in Table 3.2:
$[m(K^{\circ *})- m(K^{+*})]_{non-EM}$,
$\Gamma(K^* \rightarrow K\pi)/\Gamma(\rho \rightarrow 2\pi)$,
$\,m_\phi$ and
$\Gamma(\phi \rightarrow K {\overline K})/\Gamma(\rho \rightarrow 2\pi)$.
\begin{table}
\centerline{
\begin{tabular}{|c|c|c|c|c|}
\hline
$x$ & $(K^{\circ *}-K^{+*})_{non-EM}$ & $\Gamma(\rho)/\Gamma(K^*)$ &
$\Gamma(\rho)/\Gamma(\phi)$  & $m(\phi)$ \\
\hline
15 & $2.04\>{\rm {MeV}}$ & 5.44 & 124 & $1.01\>{\rm {GeV}}$ \\
20 & $2.44\>{\rm {MeV}}$ & 4.97 & 103 & $1.02\>{\rm {GeV}}$ \\
25 & $3.05\>{\rm {MeV}}$ & 4.45 & 81 & $1.04\>{\rm {GeV}}$ \\
28 & $3.55\>{\rm {MeV}}$ & 4.12 & 67 & $1.07\>{\rm {GeV}}$ \\
30 & $3.97\>{\rm {MeV}}$ & 3.90 & 57 & $1.09\>{\rm {GeV}}$ \\
32 & $4.47\>{\rm {MeV}}$ & 3.67 & 47 & $1.13\>{\rm {GeV}}$ \\
34 & $5.09\>{\rm {MeV}}$ & 3.44 & 38 & $1.18\>{\rm {GeV}}$ \\
36 & $5.84\>{\rm {MeV}}$ & 3.21 & 28 & $1.26\>{\rm {GeV}}$ \\
38 & $6.78\>{\rm {MeV}}$ & 2.98 & 18 & $1.43\>{\rm {GeV}}$ \\
\hline
\end{tabular}}
\caption{Predictions}
\end{table}
 These quantities are
sensitive to the wave function renormalizations of the vector particles
containing strange quarks ($K^*$ and $\phi$).  First consider the
$K^{\circ *}-K^{+*}$ mass difference.  The photon exchange contribution
has been estimated to be $-$0.7 MeV [30]. The experimental value is
alternately given as 4.5$\pm$0.4 MeV (if one simply subtracts the two numbers
in the Review of Particle Properties [31]) or as 6.7$\pm$1.2 MeV
(if one considers just the ``dedicated" experiments). So we should have,
\begin{eqnarray}
[m(K^{\circ *})-m(K^{+*})]_{non-EM} \approx \cases{ 5.2\, {\rm {MeV}}
 \cr 7.4\, {\rm {MeV}} \cr}.
\end{eqnarray}
This should be equal (see the limiting forms of (3.7) and (3.8) ) in the
$SU(3)$ limit to the negative of the non-electromagnetic piece of the
$\rho-\omega$ transition mass $M_{\rho\omega}(non-EM)$, which in turn has
been found from analysis of experiment by Gasser and Leutwyler [13] to be
$-$2.65$\pm$0.20 MeV.  Clearly the $SU(3)$ prediction is very bad.  However,
agreement can be obtained in the present model if $Z_{K^*} \not=1$ in (3.7). We
see from table 3.2 that the range of $x$ between 34 and 38 is suitable for
this purpose.  This is due to the increase of $\gamma^\prime$ with increasing
$x$.  Next, consider the ratio of widths $\Gamma(\rho \rightarrow 2\pi)/
\Gamma(K^* \rightarrow K\pi)$ which experimentally is 3.0.  Eq (3.11) shows
that this ratio is also sensitive to $Z_{K^*}$.  We see from table 3.2 that,
once again, exact agreement is obtained for $x$ around 38.  Thus considering
just the properties of the $K^*$ particle one would be tempted to choose
$x=$38.  This represents (see table 3.1) a value of $y$ significantly larger
than (3.13) in magnitude but in agreement with (3.14).  Larger values of
$x$ and $y$ (compared to (3.13) ) are also favored by consideration of the
ratio $\Gamma(\rho \rightarrow 2\pi)/\Gamma(\phi \rightarrow K {\overline K})$
which is experimentally 40.3.  We see from table 3.2 that best agreement for
this ratio is obtained for $x$ around 34.  On the other hand $m_\phi$ is fit
best with a smaller value of $x$ (about 20).  While it is true that the
simplification of neglecting $\omega-\phi$ mixing has been made above, this
effect is small enough so the $m_\phi$ prediction should be reasonably
accurate.  A compromise ``best fit" with $x\approx$32 (see (3.15)) improves
three predictions remarkably, does not distort the $m_\phi$ prediction too
badly and will be seen in the next section to be needed to lead to a
good description of the $\eta-\eta^\prime$ system.

It is of interest to note that the vector meson wave function renormalizations
for the best fit $x$=32 case are $Z_\omega=Z_\rho=0.99,\,Z_{K^*}=0.84$ and
$Z_\phi=0.65$. Since $Z_\phi$, in particular~{\footnote{In the extraction
of the parameters $c_1$ and $c_2$ of (2.5) from experiment in [22]
some $\phi$ decays are involved.  We note that
in (4.5) of this reference we should now write $\Big|\epsilon/ Z_\phi
\Big|$ instead of $|\epsilon|$.  This replacement should
also be made in (4.8) so that the old formulas for $c_1$ and $c_2$ remain
unaltered. }},
represents a rather non-trivial correction one might in the future want to
investigate other possible higher order terms involving vectors and loop
diagrams.  In any event, our analysis shows that such large symmetry breaking
is required to understand the vector mesons which contain strange quarks.

We remark here that the additional symmetry breaking term:
\begin{eqnarray}
\mu^\prime \, Tr(A_{\nu}^L {\cal M}A_{\nu}^R {\cal M})
\end{eqnarray}
can help to fine tune our results.  Because it is quadratic in ${\cal M}$
it will provide a factor of $x^2$ for $m^2(\phi)$ and negligible contributions
to other vector masses. Similarly, it will mainly affect the $\phi_{\mu}
\rightarrow K{\overline K}$ decay amplitude.  There will be an extra
contribution of $-2x^2 \mu^\prime / Z^2_\phi $ to $m^2(\phi)$ and an extra
factor of $ 1 + {{4\mu^\prime(x+x^2/2)}\over{m_v^2 + 4\alpha^\prime}}$
for the $\phi \rightarrow K {\overline K}$ amplitude. The choice
$\mu^\prime = 2.4 \times 10^{-5}\,\, {\rm {GeV}}^2$ then enables us to
fit both $m^2(\phi)$ and $\Gamma(\rho)/\Gamma(\phi)$ in Table 3.2 while
still keeping $x=31.5$.  Notice that (3.17) is the analog for the vectors
of the $\lambda^{\prime 2}$ term in (2.10) for the pseudoscalars; there are
then three analogous quark-line-rule conserving but symmetry breaking
terms for each multiplet.

\vskip.5cm

\stepcounter{chapter}
\setcounter{table}{0}
\setcounter{equation}{0}
\setcounter{footnote}{3}
\leftline{\large\it 4. Physical quantities for the $\eta-\eta^\prime$ system.}
\vskip.5cm

In the simplest Lagrangian which can mock up the $U(1)$ anomaly (just the
first two terms of (2.12) included) the $\eta^\prime$ mass and the
$\eta-\eta^\prime$ mixing can be satisfactorily fit.  However the $\eta$ mass
comes out too low {\footnote {This can be seen by referring to Fig 2 of
[32].}}.
Here we will show that this problem can be remedied
if the second two terms of (2.12) as well as the OZI rule conserving symmetry
breakers in (2.10) are all taken into account.

Defining the 2-dimensional vector $\eta=\pmatrix{\eta_T \cr \eta_S \cr}$ where
$\eta_T =(\phi_{11}+\phi_{22})/\sqrt2$, $\eta_S = \phi_{33}$, we collect the
quadratic terms in the effective Lagrangian as
\begin{eqnarray}
-{1\over2}\partial_\mu\eta^\top K\partial_\mu\eta -{1\over2}\eta^\top P\eta .
\end{eqnarray}
Here the ``kinetic" matrix is
\begin{eqnarray}
K_{TT} &=& 1 + {4\over{F_{\pi}^2}}({\alpha^\prime\over{\tilde
g}^2}-4\beta^\prime)+  {{16\,n}\over{F_\pi^2}}, \nonumber \\
K_{SS} &=& 1+{{4x}\over{F_{\pi}^2}}({\alpha^\prime\over{\tilde
g}^2}-4\beta^\prime)+ {{8\,n}\over{F_\pi^2}} , \nonumber \\
K_{TS}&=&K_{ST} = {{8\sqrt2\,n}\over{F_\pi^2}},
\end{eqnarray}
while the ``potential" matrix is
\begin{eqnarray}
P_{TT}&=&{1\over{F_\pi^2}}\big[8\delta^\prime+{8\over3}\kappa\epsilon^\prime +
{\kappa\over9} +16\kappa\epsilon^{\prime2}+32\lambda^{\prime2}(1+y^2)\big],
\nonumber \\
P_{SS}&=&{1\over{F_\pi^2}}\big[8x\delta^\prime+{4\over3}x\kappa\epsilon^\prime
+{\kappa\over{18}}+8\kappa x^2\epsilon^{\prime2}+32 x^2\lambda^{\prime2}\big] ,
\nonumber \\
P_{TS}&=&P_{ST}= {{\sqrt2}\over{F_\pi^2}}\,\big[{2\over3}\kappa\epsilon^\prime
(1+x)+{\kappa\over{18}}+8\kappa x\epsilon^{\prime2}\big].
\end{eqnarray}
The ``kinetic" matrix is diagonalized by
\begin{eqnarray}
R^{-1}(\theta_1)K\,R(\theta_1)={\hat K}=\pmatrix{{\hat K}_1 & 0 \cr
0 & {\hat K}_2 \cr} ,
\end{eqnarray}
where
\begin{eqnarray}
R(\theta_i)=\pmatrix{\cos\theta_i & \sin\theta_i \cr -\sin\theta_i &
\cos\theta_i \cr} ,
\end{eqnarray}
and ${\hat K}_{1,2} = {1\over2}[(K_{TT}+K_{SS}) \mp{\sqrt{(K_{TT}-K_{SS})^2+
4K_{TS}^2}}\,]$.  The angle $\theta_1$ is obtained from
\begin{eqnarray}
\tan 2\theta_1={{2K_{TS}}\over{K_{SS}-K_{TT}}} .
\end{eqnarray}
The {\it{effective}} potential matrix $P^\prime$ is then
\begin{eqnarray}
P^\prime={\hat K}^{-1/2}R^{-1}(\theta_1)P\,R(\theta_1){\hat K}^{-1/2}.
\end{eqnarray}
This yields the {\it{physical}} $\eta$ and $\eta^\prime$ squared masses
as the eigenvalues of (4.7):
\begin{eqnarray}
m^2(\eta,\eta^\prime)={1\over2}\big[(P_{TT}^\prime+P_{SS}^\prime)\mp
{\sqrt{(P_{TT}^\prime-P_{SS}^\prime)^2+4P^{\prime2}_{TS}}}\big].
\end{eqnarray}
The ``bare" column vector is finally expanded in terms of the {\it
{physical}} fields $\eta_p=\pmatrix{ \eta  \cr \eta^\prime \cr}$ as
\begin{eqnarray}
 \eta=R(\theta_1){\hat K}^{-1/2}R(\theta_2)\eta_p ,
\end{eqnarray}
where $\theta_1$ is obtained from (4.6) while $\theta_2$ is determined from
\begin{eqnarray}
 \tan\,2\theta_2 = {{2P^{\prime}_{TS}}\over{P^{\prime}_{SS}-P^{\prime}_{TT}}}.
\end{eqnarray}
Note that the transformation (4.9) is, in general, non-orthogonal.  For
comparison, the mixing convention in the octet-singlet basis is
\begin{eqnarray}
\pmatrix{\eta_8 \cr \eta_1 \cr }=R(\theta)\,\pmatrix{\eta \cr \eta^\prime \cr }
\end{eqnarray}
and, to the extent that ${\hat K}$ may be approximated by the unit matrix,
\begin{eqnarray}
\theta \approx \theta_1+\theta_2 - 54.7^{\circ} .
\end{eqnarray}

The standard source of information, which we shall also employ here, about
mixing in the $\eta-\eta^\prime$ system is comparison of the
$\pi^\circ \rightarrow 2\gamma$, $\eta^\prime \rightarrow 2\gamma $ and
$\eta \rightarrow 2\gamma $ decays.  It is well known that
$\pi^\circ \rightarrow 2\gamma$ is well-described by ``gauging" the first term
of (2.5). Thus it is sufficient to consider ratios of the other rates to the
$\pi^\circ$ rate.  We find
\begin{eqnarray}
{{\Gamma(\pi^\circ \rightarrow 2\gamma)}\over{m^3_\pi}}:
{{\Gamma(\eta \rightarrow 2\gamma)}\over{m^3_\eta}}:
{{\Gamma(\eta^\prime \rightarrow 2\gamma)}\over{m^{3}_{\eta^\prime}}} \,=\,
{9\over2}:(a_1)^2:(a_2)^2,
\end{eqnarray}
wherein
\begin{eqnarray}
a_1 &=& {5\over{\sqrt2}}\left({{\cos\theta_1\,\cos\theta_2}\over{{\hat K}^{1/2}
_1}} - {{\sin\theta_1\,\sin\theta_2}\over{{\hat K}^{1/2}_2}}\right) - \left(
{{\sin\theta_1\,\cos\theta_2}\over{{\hat K}^{1/2}_1 }} +
{{\sin\theta_2\,\cos\theta_1}\over{{\hat K}^{1/2}_2}} \right)\nonumber \\
a_2 &=& {5\over{\sqrt2}}\left({{\cos\theta_1\,\sin\theta_2}\over{{\hat K}^{1/2}
_1}} + {{\cos\theta_2\,\sin\theta_1}\over{{\hat K}^{1/2}_2}}\right) + \left(
{{\cos\theta_1\,\cos\theta_2}\over{{\hat K}^{1/2}_2 }} -
{{\sin\theta_1\,\sin\theta_2}\over{{\hat K}^{1/2}_1}} \right) .
\end{eqnarray}
Experimentally, $(a_1)^2=3.98\pm 0.60$ and $(a_2)^2=7.20\pm 1.40$.

We may now try fit our parameters to these experimental data. Adopting the
compromise fit discussed in Sec.3 with $x=31.5$, we have three new unknown
parameters: $\kappa, \epsilon^\prime$ and $n$ (see (2.12) ).  On the other hand
there are four experimental quantities to be fit: $m_{\eta}, m_{\eta^\prime}$,
and  the $\eta-\eta^\prime$ mixing information contained in $(a_1)^2$ and
$(a_2)^2$ of (4.13).  Conducting a numerical scan of the $\kappa,
\epsilon^\prime$ and $n$ parameter space we find a best fit for
\begin{eqnarray}
\kappa=0.105 \,{\rm {GeV}}^4,\quad n= 1.40\times10^{-4} \,{\rm {GeV}}^2, \quad
\epsilon^\prime=1.59\times10^{-4}.
\end{eqnarray}
With these parameters the physical quantities are
\begin{eqnarray}
m(\eta)=549\,{\rm {MeV}},\>m(\eta^\prime)=959\,{\rm {MeV}},\>
(a_1)^2=3.98,\> (a_2)^2=7.19,
\end{eqnarray}
in agreement with experiment.  Furthermore, the quantities describing the
non-orthogonal $\eta-\eta^\prime$ matrix in (4.9) are
\begin{eqnarray}
\theta_1=7.44^\circ,\> \theta_2=34.7^\circ,\>{\hat K}_1^{1/2}=1.07,\>
{\hat K}_2^{1/2} =1.36.
\end{eqnarray}
This would correspond, via (4.12), to a usual $\eta-\eta^\prime$ mixing angle
of about $-12.6^\circ$.  However, because ${\hat K} \not= 1$ this comparison
is just suggestive.

We have found that it is difficult to achieve a fit for the experimental
quantities in the $\eta-\eta^\prime$ system for $x$ significantly different
from the value used above.  It is also amusing to note that the $\lambda^{
\prime2}$ symmetry breaker in (2.10) plays an important role in enabling us
to get a good fit in the first place.  Without the $\lambda^{\prime2}$ term
present the only reasonable fit was found with very large and negative
$\epsilon^\prime$ which would give the wrong sign for the neutron-proton
mass difference, as computed in this model.
\vskip.3cm

\stepcounter{chapter}
\setcounter{equation}{0}
\setcounter{table}{0}
\setcounter{footnote}{4}
\leftline{\large\it 5. The neutron proton mass difference}
\vskip.3cm

In the preceding sections we discussed the effective meson
action with emphasis on the symmetry breaking parts, especially those
which are isospin non-invariant. In this section we will use
this effective mesonic model to obtain information
about properties of baryons. In mesonic theories like the one under
consideration baryons emerge as soliton solutions.

In order to explicitly test the isospin non-invariant terms we
will evaluate the non-electromagnetic contribution to the neutron
proton mass difference:
\begin{eqnarray}
\Delta=\big(M_{\rm neutron}-M_{\rm proton}\big)_{\rm non-EM}.
\end{eqnarray}
Experimentally the neutron proton mass difference is [31]
$1.29 MeV$. However the
electromagnetic interaction (photon exchange) also
 contributes to the neutron proton mass
difference. Using experimental electromagnetic form factors this
contribution is estimated as [13]:
$$\Delta_{EM}=(-0.76\pm0.30)MeV.$$
The negative sign agrees with $\Delta_{EM}$ being
dominated by the Coulomb repulsion. Thus we may extract:
\begin{eqnarray}
\Delta=(2.05\pm0.30)MeV.
\end{eqnarray}

\leftline{\it a. Description of the approach}
\vskip.3cm

It has been demonstrated previously[1] that vector mesons are a necessary
 ingredient in  an effective meson theory in order for it to
explain the  neutron proton mass difference reasonably well. In
ref.[1]  however, only the $SU(2)\times U(1)$ version of the model
was properly calculated while the $SU(3)$ part was just  roughly estimated.
It is therefore highly desirable to extend the treatment to $SU(3)$.

In the first step the static soliton with unit baryon number is
constructed\footnote{More details on the notation of this section may
be found in refs [17].}.
 The appropriate classical ans\"atze are:
\begin{eqnarray}
\xi_\pi({\bf r})=\pmatrix {exp\big(i{\bf \hat r}\cdot
\mbox{\boldmath $\tau$} F(r)/2\big) & 0 \cr 0 & 1 \cr},\quad
\omega_0={{\omega(r)}\over{2 \sqrt2\tilde g}},\quad
\rho_{i,a}={{G(r)}\over{\sqrt2 \tilde gr}}\epsilon_{ija}\hat r_j.
\end{eqnarray}
Substituting (5.3) into the Lagrangian ${\cal L}$ yields the
classical  mass $M_{\rm cl}=-\int d^3r {\cal L}$.  The isospin
breaking terms do not actually  contribute to $M_{\rm cl}$. An analytic
expression for $M_{\rm cl}$ may be found in refs.[17]. Extremizing
$M_{\rm cl}$ yields second order non-linear differential equations
for $F(r), G(r)$  and $\omega(r)$. The solutions to these equations
are characterized by their topological charge which is identified
with the baryon number. The solution of the baryon number one sector
has been extensively  discussed in the literature[17-20,22].
The static soliton (5.3) is invariant under ``grand spin"
${\bf G} = {\bf J} + {\bf I}$ transformations but not under
spin (${\bf J}$) and isospin (${\bf I}$) transformations
separately. Therefore the static soliton carries neither
good spin nor good isospin quantum numbers. In $SU(2)$ the projection
onto good quantum numbers is carried out by introducing  time
dependent collective coordinates for the zero modes of the theory.
In $SU(2)$ these are the spatial and isospatial rotations. These two
transformations are actually equivalent due to the above mentioned
grand spin symmetry. Thus in $SU(2)$ the spectrum contains only
baryons with equal spin and isospin, {\it e.g.} N,$\Delta$.

Unfortunately the extension of soliton models to include
strange fields is non-trivial due to the presence of $SU(3)$ breaking,
{\it i.e.} flavor transformations are not real zero modes.
However, we will consider $SU(3)$ to still be an approximate symmetry and
therefore shall introduce collective coordinates for the whole flavor group.
This approach allows us to  easily make  contact with ordinary baryon
phenomenology. Furthermore it has been demonstrated in ref.[17] that the
collective approach describes static properties of the low-lying
${1\over2}^+$ and ${3\over2}^+$ baryons reasonably well.
An alternative point of view that $SU(3)$ symmetry
is strongly broken leads to the bound state approach[33]. In this
approach a kaonic bound state is constructed in the background
field of the static soliton. The corresponding bound state energy
eigenvalue determines the mass splittings of baryons with
different hypercharge since they are characterized by different
occupation numbers of this bound state.
Numerically the two (somewhat different) approaches yield comparable
results for the mass differences of baryons with different hypercharge
and spin.

The collective coordinates $A(t)\in SU(3)$ are introduced by
flavor-rotating (5.3):
\begin{eqnarray}
\xi({\bf r},t)=A(t)\xi_k\xi_\pi({\bf r})\xi_kA^\dagger(t)
\end{eqnarray}
and
\begin{eqnarray}
\frac{1}{\sqrt2}\rho_\mu({\bf r},t)=
A(t)\pmatrix{\rho^\pi_\mu+\omega_\mu &K^*_\mu \cr
{K^{*\dag}_\mu}& 0\cr}A^\dagger(t).
\end{eqnarray}
Here we have also allowed for a more general field configuration
parametrized by $\xi_k$ and $K_{\mu}^*$. We will see shortly how these
additional fields get excited. The time dependence of the
collective coordinates is made most transparent by the introduction
of ``angular velocities", $\Omega_a$ (a=1,...,8):
\begin{eqnarray}
A^{\dag}{\dot A}={i\over2}{\sum^8_{a=1}}{\lambda_a}{\Omega_a}=
i\pmatrix{{\Omega_\pi}+{\Omega_\eta}&\Omega_K \cr
{\Omega_K^{\dag}}&-2{\Omega_\eta}\cr}
\end{eqnarray}
wherein the $\lambda_a,\quad (a=1,..,8)$ denote the $SU(3)$ Gell-Mann
matrices.

Obviously the vector meson  terms in (2.4) as well as the
``anomalous" part of the action (2.5) contain expressions which are
linear in the time derivative. These expressions provide source terms
linear in the angular velocity for the fields which vanish classically. Thus
additional fields are excited by the collective rotation. Clearly they
are linear in the angular velocities. The most general ansatz for
the non-strange vector mesons excited by the isospin rotation is[17,19-20]
\begin{eqnarray}
\rho_0^\pi ={1\over {2\sqrt2 \tilde g}}[\xi_1(r) {{\bf \Omega}} +
\xi_2(r)({{\bf \hat r}}\cdot {{\bf \Omega}}){{\bf\hat r}}]
\cdot {\mbox{\boldmath $\tau$}},\quad \omega_i
={{\Phi(r)}\over {2 \sqrt2\tilde g}}\epsilon_{ijk}\Omega_j\hat r_k.
\end{eqnarray}
The pseudoscalar nonet contains components which are excited by the isospin
rotation[20] as well as by  rotation into the strange directions[19].
Parametrizing:
\begin{eqnarray}
\xi_k=e^{iz},\quad z=\pmatrix{\eta_T &K\cr K^\dagger&\eta_S\cr}
\end{eqnarray}
suitable ans\"atze are $\eta_T=\frac{1}{4}(\chi(r)+\chi_8(r))
{\bf \hat r}\cdot{\bf \Omega}$,
$\eta_S=\frac{1}{4}(\chi(r)-2\chi_8(r))
{\bf \hat r}\cdot{\bf \Omega}$ and $K=W(r){\bf \hat r}\cdot
{\mbox{\boldmath $\tau$}}\Omega_K$. For later calculations it
is convenient to introduce also an ansatz for the glueball field
$G=g(r){\bf \hat r}\cdot{\bf \Omega}$ instead of eliminating
$G$ from the beginning via  (2.13). From parity
and isospin covariance we also get the following ans\"atze for the $K^*$
isospinor fields:
\begin{eqnarray}
K^*_0={{S(r)}\over{\sqrt2 \tilde g}}\Omega_K,\quad
K^*_i={1\over {2 \sqrt2 \tilde g}}[iE(r)\hat r_i +
{D(r)\over r}\epsilon_{ijk}\hat r_j \tau_k]\Omega_K.
\end{eqnarray}
Substituting (5.3-5,7-9) into the action and expanding up to second
order in the angular velocities as well as linear order in isospin
breaking yields the collective Lagrangian:
\begin{eqnarray}
L=&-M_{\rm cl}+{1\over2}\alpha^2\sum_{i=1}^3\Omega_i^2+
{1\over2}\beta^2\sum_{\alpha=4}^7\Omega_\alpha^2-
{{\sqrt3}\over2}\Omega_8+\alpha_1\sum_{i=1}^3D_{8i}\Omega_i
+\beta_1\sum_{\alpha=4}^7D_{8\alpha}\Omega_\alpha \nonumber \\
&-{1\over2}\gamma(1-D_{88})-{1\over2}\gamma_S(1-D_{88}^2)
-{1\over2}\gamma_T\sum_{i=1}^3D_{8i}D_{8i}
-{1\over2}\gamma_{TS}\sum_{\alpha=4}^7D_{8\alpha}D_{8\alpha} \nonumber \\
&-\Gamma_3 D_{38}-\Delta_3\sum_{i=1}^3(D_{3i}D_{8i}+D_{38}D_{88})
+\alpha_3\sum_{i=1}^3D_{3i}\Omega_i
+\beta_3\sum_{\alpha=4}^7D_{3\alpha}\Omega_\alpha,
\end{eqnarray}
where the $D_{ij}$ denote the matrices ${1\over2}Tr(\lambda_iA\lambda_j
A^\dagger)$ of the $SU(3)$ adjoint representation.
The moments of inertia $\alpha^2$ and $\beta^2$ are functionals of
the radial functions $(\xi_1,\xi_2,\Phi,\chi,\chi_8)$ and
$(W,S,E,D)$, respectively. Varying the moments of inertia with
respect to these fields yields linear inhomogeneous second order
differential equations with the classical fields $F,G$ and
$\omega$ as source terms. Analytic expressions for the moments of
inertia as well as the symmetry breaking parameters $\gamma,\alpha_1$
and $\beta_1$ may be found in ref.[17]. Here we only wish to explain
the mechanism which excites $\chi,\chi_8$ and $g$. The non-strange
combination $\chi+\chi_8$ is excited by the terms proportional to
$\epsilon_{\mu\nu\alpha\beta}$ in (2.14) exactly as explained in
refs.[1,20]. The glueball field and the strange part of the $\eta$ fields
are subsequently induced via the equation of motion (2.13).

The additional symmetry breaking parameters
$\gamma_S,\gamma_T$ and $\gamma_{TS}$ are solely due to the
$(\lambda^\prime)^2$ term in (2.10):
\begin{eqnarray}
\gamma_S & =-\gamma_T=-{{32\pi}\over9}(\lambda^\prime)^2
(1-x)^2\int dr r^2 (1-{\rm cos}2F),\nonumber \\
\gamma_{TS} & ={{32\pi}\over3}(\lambda^\prime)^2
(1-x)^2\int dr r^2 (1-{\rm cos}F).
\end{eqnarray}
The isospin symmetry breaking parameters $\Gamma_3,\Delta_3,\alpha_3$
and $\beta_3$ are needed for the evaluation of the
neutron proton mass difference and we  present the somewhat
lengthy analytic expressions in an appendix.
Here we just wish to mention that $\Gamma_3$ and $\Delta_3$ contain
only classical fields while $\alpha_3$ and $\beta_3$ contain expressions
linear in the excitations (5.7-8) as well.

The Lagrangian (5.10) is quantized canonically by introducing $SU(3)$
right generators, $R_a\  (a=1,..,8)$ via:
\begin{eqnarray}
R_a=-{{\partial L}\over{\partial\Omega_a}}=\cases{
-(\alpha^2\Omega_a+\alpha_1D_{8a}+\alpha_3D_{3a}),&a=1,2,3 \cr
-(\beta^2\Omega_a+\beta_1D_{8a}+\beta_3D_{3a}),&a=4,..,7 \cr
{1\over2}\sqrt3,&a=8.\cr}
\end{eqnarray}
We separate the isospin breaking part of the Hamiltonian:
\begin{eqnarray}
H=-\sum_{a=1}^8R_a\Omega_a-L=H_{I=0}+H_{I=1}
\end{eqnarray}
The isospin symmetric part $H_{I=0}$ may be diagonalized exactly by
generalizing the approach of Yabu and Ando[16]. This yields
the energy formula:
\begin{eqnarray}
E_{I=0}=M_{cl}+{1\over2}\big({1\over{\alpha^2}}-{1\over{\beta^2}}\big)J(J+1)
-{3\over{8\beta^2}}+{1\over{2\beta^2}}\epsilon_{SB},
\end{eqnarray}
where $\epsilon_{SB}$ is the eigenvalue of
\begin{eqnarray}
C_2&+\beta^2\gamma(1-D_{88})+\beta^2{{\alpha_1}\over{\alpha^2}}
\sum_{i=1}^3 D_{8i}(2R_i+\alpha_1D_{8i})+
\beta_1\sum_{\alpha=4}^7D_{8\alpha}(2R_\alpha+\beta_1D_{8\alpha})
\nonumber \\
&+\beta^2\gamma_S(1-D_{88}^2)+\beta^2\gamma_T\sum_{i=1}^3D_{8i}D_{8i}
+\beta^2\gamma_{TS}\sum_{\alpha=4}^7D_{8\alpha}D_{8\alpha},
\end{eqnarray}
and $C_2=\sum_{a=1}^8R_a^2$ denotes the quadratic Casimir operator of
$SU(3)$. $\epsilon_{SB}$ is obtained, of course, numerically, using
differential
operator realizations of the $R_a$ [34].
The isospin breaking part is obtained to be:
\begin{eqnarray}
H_{I=1}&=&\Gamma_3D_{38}+
\Delta_3\sum_{i=1}^3(D_{3i}D_{8i}+D_{38}D_{88})
\nonumber \\
&&+{{\alpha_3}\over{\alpha^2}}\sum_{i=1}^3D_{3i}(R_i+\alpha_1D_{8i})
+{{\beta_3}\over{\beta^2}}\sum_{\alpha=4}^7
D_{3\alpha}(R_\alpha+\beta_1D_{8\alpha}).
\end{eqnarray}
In the evaluation of the Hamiltonian only  terms linear in the isospin
breaking have been retained. Finally we have the non-electromagnetic
contribution to the neutron proton mass
difference:
\begin{eqnarray}
\Delta&=&\langle n|H_{I=1}|n \rangle
-\langle p|H_{I=1}|p \rangle
\nonumber \\
&=&-2\Gamma_3\langle p|D_{38}|p \rangle
-2\Delta_3\langle p|\sum_{i=1}^3(D_{3i}D_{8i}+D_{38}D_{88})|p \rangle
\nonumber \\
&&-2{{\alpha_3}\over{\alpha^2}}\sum_{i=1}^3
\langle p|D_{3i}(R_i+\alpha_1D_{8i})|p \rangle
-2{{\beta_3}\over{\beta^2}}\sum_{\alpha=4}^7
\langle p|D_{3\alpha}(R_\alpha+\beta_1D_{8\alpha})
|p \rangle,
\end{eqnarray}
where $|p \rangle$ denotes the exact proton eigenstate of $H_{I=0}$. In
writing (5.17)
we have made use of the flavor transformation properties of the
D-functions.

Although it is obvious, we would like to stress that in this treatment
the neutron proton mass splitting is not obtained as the difference of
two large numbers. On the contrary, the leading operator in the isospin
breaking($H_{I=1}$) has been extracted and its expectation value is identified
with the mass difference.

  \vskip.3cm
\leftline{\it b. Numerical results}
\vskip.3cm

We now present our numerical results for the
neutron proton mass difference. To proceed we have to fix the parameters
of the anomalous part of the action, (2.5). Previously [22] it has been shown
 that
\begin{eqnarray}
\tilde{h}&=&4(2c_1-c_2/\tilde g-c_3/4\tilde g^2) \approx 0.4 \nonumber \\
\tilde{g}_{VV\phi}&=&4c_2 \approx 1.9
\end{eqnarray}
fit reasonably well the decay processes $\omega\rightarrow3\pi$ and
$\omega\rightarrow\rho\pi$. $\tilde{h}$ and $\tilde{g}_{VV\phi}$ are
allowed to vary {\footnote{Hans Walliser has pointed out to us that the
experimental measurements for the relevant $\omega$ and $\phi$ partial
decay widths have changed from the 1986 to the 1992 ``Review of Particle
Properties". Continuing to use (4.4) [see footnote 3 above], (4.7)
and (4.9) of [22] now gives us $|\epsilon|/Z_\phi = 0.059 \pm 0.005$
and central values $\tilde {g}_{VV\phi} = 1.81, \, \tilde{h}=0.38$.
These are substantially similar to the older values.}}
in the range $\tilde h=-0.15,..,0.7$ and
$\tilde g_{VV\phi}=1.3,..,2.2$ subject to the constraint
$\vert\tilde g_{VV\phi}-\tilde h\vert\approx 1.5$ due to uncertainties
in the determination of the $\omega-\phi$ mixing angle. The third parameter
could not be fixed in the meson sector; however, it was argued that
$\kappa=c_3/2\tilde gc_2\approx1$ from studies of baryon properties. We will
adopt this value together with (5.18).

In table 5.1 we present the results for the isospin breaking parameters
$\Gamma_3, \Delta_{3}, \alpha_3$ and $\beta_3$ as functions
of $x$. Obviously $\Delta_{3}$ is negligible and we will omit
it from now on.
\begin{table}
\centerline{
\begin{tabular}{|c|c|c|c|c|}
   \hline
$x$ & $\Gamma_3(MeV)$ & $\Delta_{3}(MeV)$ & $\alpha_3$ & $\beta_3$  \\
\hline
$25.0$&-15.32&0.08&-0.0046&-0.0171 \\
\hline
$28.0$&-15.60&0.05&-0.0049&-0.0178 \\
\hline
$30.0$&-16.08&0.04&-0.0052&-0.0185 \\
\hline
$31.5$&-16.45&0.03&-0.0055&-0.0189 \\
\hline
$34.0$&-16.89&0.02&-0.0060&-0.0195 \\
\hline
$38.0$&-17.89&0.00&-0.0070&-0.0205 \\
\hline
\end{tabular}}
\caption{The isospin breaking parameters $\Gamma_3, \Delta_{3}, \alpha_3$
and $\beta_3$ as functions of $x$. The input parameters are
according to table 3.1. For parameters of the anomalous sector we
take $\tilde{h}=0.4, \tilde g_{VV\phi}=1.9, \kappa=1.0$.}
\end{table}

In table 5.2 we display the relevant matrix elements and the prediction
for the non-electromagnetic contribution to the neutron proton mass difference,
$\Delta$ in (5.3).

\begin{table}
\centerline{
\begin{tabular}{|c|c|c|c|c|}
   \hline
$x$ & $-2 \langle D_{38}\rangle_p$ &
$M_1 (MeV) $ &
$M_2 (MeV) $ &
$M_n-M_p (MeV)$ \\
\hline
$25.0$&-0.017&-174.1&-31.0&1.60 \\
\hline
$28.0$&-0.020&-180.0&-31.2&1.75 \\
\hline
$30.0$&-0.021&-183.5&-31.0&1.88 \\
\hline
$31.5$&-0.023&-185.1&-30.6&1.97 \\
\hline
$34.0$&-0.025&-188.6&-29.7&2.12 \\
\hline
$38.0$&-0.028&-193.0&-27.3&2.41 \\
\hline
\end{tabular}}
\caption{The relevant matrix elements
$M_1={{-2}\over{\alpha^2}}\sum_{i=1}^3
\langle D_{3i}(R_i+\alpha_1D_{8i})\rangle_p $ and
$M_2={{-2}\over{\beta^2}}\sum_{\alpha=4}^7
\langle D_{3\alpha}(R_\alpha+\beta_1D_{8\alpha})\rangle_p $
to evaluate the neutron proton mass difference as functions
of $x$. Input parameters as in table 5.1}
\end{table}
Note that the $\alpha_3$ term  dominates and contributes more than
$50\%$ of the neutron proton mass difference. This is
expected since the $\alpha_3$ term represents essentially the pure two flavor
contribution. As previously discussed [1] the use
of the exact wave functions of the $SU(3)$ Hamiltonian rather than the
``unperturbed" wave function drastically suppresses the matrix elements of
the $\Gamma_3$ piece.
Clearly, reasonable agreement with the result extracted from experimental
data (5.2) is obtained for $x\ge 30$. Especially the prediction for the
neutron mass difference for the favored value in the meson
sector $x=31.5$ turns out to almost coincide with the central value
of the experimental data.

Based on the discussion in ref.[17]  one might however object that for
$\tilde h=0.4$ and $\tilde g_{VV\Phi}=1.9$, the $SU(2)$ moment of
inertia $\alpha^2=4.52GeV^{-1}$ is somewhat too small, predicting too
large mass splittings between the $\frac{1}{2}^+$ and $\frac{3}{2}^+$
baryons. A small variation to $\tilde h=0.36$, $\tilde g_{VV\Phi}=1.88$
and $x=28$ yields $\alpha^2=5.00GeV^{-1}$.  The corresponding mass
differences are displayed in table 5.3 and are found to agree reasonably
well with the experimental data.  Actually the $SU(3)$ mass differences
come out slightly worse than previously since the effect of
$\alpha^\prime$ is lowered.  This is due to the fact that large, positive
$\alpha_1$ and $\beta_1$ terms in  (5.10) mitigate the effects of a large
collective symmetry breaking parameter $\gamma$. In the previous fit
(without the $\gamma^\prime$ type term) $\alpha^\prime$ was negative and
significantly larger in magnitude than in the present case (cf. table 3.1).
Noting\footnote {Correct charge normalization requires that we only include
the contributions of the classical fields (5.3) to $\alpha_1$ and $\beta_1$.}
that $\alpha_1$ is essentially proportional to $(1-x)\alpha^\prime$ we
may understand why the $SU(3)$ mass differences are increased.
For this parameter set we  obtain:
\begin{eqnarray}
\Delta=1.77MeV \>,
\end{eqnarray}
which is slightly lower than the central value $2.05 MeV$ of the experimental
data for $\Delta$ but still within the error bars. The dominant contribution
to (5.19) stems from the $\alpha_3$ term: $0.93MeV$. Previously[1] we had found
that this contribution was almost entirely due to the $\eta$ excitation.
This is now somewhat modified since firstly in ref. [1] only the
unphysical $\eta_T$ was present and its mass of $550MeV$ was assumed.
Employing the treatment of section 4 in the soliton sector as well decreases
the effect of the $\eta$ fields somewhat because they are suppressed
by a larger mass. Secondly the $\gamma^\prime$ term which was not present
in the approach of ref.[1] contributes about $30\%$ to $\alpha_3$. The
strange excitations contribute about $0.54MeV$ to the mass difference.
The contribution of the $\Gamma_3$ term ($0.29MeV$) is strongly suppressed
since the matrix element of $D_{38}$ is much smaller than its
$SU(3)$ symmetric value when the exact eigenstates of $H_{I=0}$ are
used.
Comparing  (5.19) with the results obtained for the central
values (5.18) we recognize a small change for the worse with the
baryon best fit parameters. However, the better agreement of
$\Delta$ for the set (5.18) is merely due to the smaller non-strange
moment of inertia so we would expect the baryon best fit
set to be more reliable for predicting baryon observables.

\begin{table}
\centerline{
\begin{tabular}{|c|c|c|c|c|c|c|c|}
   \hline
&$\Lambda$&$\Sigma$&$\Xi$&$\Delta$&$\Sigma^*$&$\Xi^*$&$\Omega$  \\
\hline
Fit&168.&263.&404.&327.&470.&617.&766. \\
\hline
(5.18)&144.&237.&350.&360.&480.&602.&723. \\
\hline
Expt.&177.&254.&379.&293.&446.&591.&733. \\
\hline
\end{tabular}}
\caption{Best fit to the mass differences of the
$\frac{1}{2}^+$ and $\frac{3}{2}^+$
baryons with respect to the nucleon for the parameter set $\tilde h=0.36$,
$\tilde g_{VV\Phi}=1.88$, $\kappa=1.0$ and $x=28$
(see also table 3.1). Also listed are the predicted mass differences
for the central values for $\tilde h$ and $\tilde g_{VV\Phi}$ with
$\kappa=1.0$ and $x=31.5$ (3.15).}
\end{table}

In the next section we will discuss a small extension of the model Lagrangian
which enables us to fine tune the  predicted neutron proton mass
difference by adding a term which may be interpreted as providing
the gluonic contribution to the proton spin.
\vskip.3cm

\stepcounter{chapter}
\setcounter{table}{0}
\setcounter{equation}{0}
\setcounter{footnote}{7}
\leftline{\large\it 6. Two component description of the proton spin puzzle}
\vskip.3cm

In this section we will describe the implications of the preceding
considerations on the so called proton spin puzzle which actually
refers to the surprisingly small matrix element of the axial singlet
current between proton states. In a na\"\i ve quark model this matrix
element measures twice the proton spin. As indicated previously[3] the
neutron proton mass difference may be used to determine the gluonic
contribution to this matrix element. In ref.[3] only an estimate of
this quantity was made. Here we will present the complete calculation
since all the necessary tools have been provided in the preceding
sections  and we may therefore disentangle the two components of
the axial singlet matrix element: matter and gluon contribution [11].

The axial singlet current, $J_\mu^5$ may most easily be obtained from
\begin{eqnarray}
J_\mu^5=-2\frac{\partial {\cal L}}{\partial_\mu \chi}=
\sqrt3 F_\pi\partial_\mu\eta^\prime+s\tilde J_\mu^5
\end{eqnarray}
with $\chi$ being defined after eqn. (2.13). It is important to note that
the second part,
\begin{eqnarray}
\tilde J_\mu^5 & = & \epsilon_{\mu\nu\rho\sigma}
{\rm Tr} \Big\{\frac{i}{2}\big(\frac{\gamma_1}{3}+\frac{\gamma_2}{2}\big)
\tilde p_\nu\tilde p_\rho V_\sigma-\sqrt2\gamma_2\tilde g
F_{\nu\rho}V_\sigma
-4i\tilde g^2(\gamma_2+2\gamma_3)V_\nu V_\rho V_\sigma\Big\}
\end{eqnarray}
is not a pure gradient and has therefore a non-vanishing
matrix element at zero momentum transfer in contrast to the first part.
In eqn. (6.2) we have used $\frac{1}{\sqrt2}V_\mu=\rho_\mu-\frac{i}{2g}v_\mu$
as
 well as
$\tilde p_\mu(v_\mu)=\tilde U^{1/2\dagger}\partial_\mu \tilde U^{1/2}\pm
\partial_\mu\tilde U^{1/2}\tilde U^{1/2\dagger}$.  $\tilde J_\mu^5$
may be considered as the short distance part of the axial singlet
current since it is dominated by the vector mesons and is not present
in models with pseudoscalars only. The dimensionless
constant $s$ has been introduced by the replacement in the Lagrangian:
\begin{eqnarray}
\partial_\mu\chi\tilde J_\mu^5\longrightarrow
s\partial_\mu\chi\tilde J_\mu^5
\end{eqnarray}
to allow for a deviation from the nonet form for the pseudoscalar
fields in the Lagrangian. The
nonet form was originally introduced to satisfy the OZI rule; however
such an approach seems somewhat debatable for the $\eta^\prime$.

Defining axial form factors of the nucleon for the flavor $l=u,d,s$ via
\begin{eqnarray}
{{\sqrt{p_0p_0^\prime}}\over{M_P}}&&<P(\vec p\ ^\prime)|
{\overline q}_l\gamma_\mu\gamma_5q_l|P(\vec p)>= \nonumber \\
&& {\overline u}(\vec p\ ^\prime)\big[\gamma_\mu\gamma_5H_l(q^2)+
{{i q_\mu}\over{2M_P}}\gamma_5\tilde H_l(q^2)\big]u(\vec p),
\quad {\rm with}\  q_\mu=p_\mu-p_\mu^\prime.
\end{eqnarray}
the relevant quantity for the axial singlet current is
\begin{eqnarray}
H(q^2)=\sum_{l=1}^3H_l(q^2).
\end{eqnarray}
Obviously the first term in (6.1) only contributes to the induced
form factor $\tilde H(q^2)=\sum_{l=1}^3\tilde H_l(q^2)$. In the
original Skyrme model (without vector mesons) $\tilde J_\mu^5=0$
and therefore also $H(0)=0$, which has been considered as a
success[7] of the Skyrme model since it nicely describes the
results of the EMC experiment[4].

In order to introduce the two component mechanism [11] it is necessary [3]
to allow the pseudoscalar gluon field, $G$ to couple to $\tilde J_\mu^5$
via the chirally invariant expression:
\begin{eqnarray}
\frac{2t}{\kappa}\partial_\mu G\tilde J_\mu^5
\end{eqnarray}
wherein $t$ is a new dimensionless parameter. This additional term
changes the equation of motion for the glueball field (2.13) to
\begin{eqnarray}
G = \frac{\kappa}{2\sqrt3 F_\pi}\eta^\prime
+t\partial_\mu\tilde J_\mu^5.
\end{eqnarray}
We furthermore have the $U_A(1)$ anomaly equation which is
equivalent to the equation of motion for $\eta^\prime$ after
eliminating the glueball field:
\begin{eqnarray}
(-\partial^2+\frac{\kappa}{6F_\pi^2})\eta^\prime=
\frac{s-t}{\sqrt3 F_\pi}\partial_\mu\tilde J_\mu^5
\end{eqnarray}
In both equations (6.7) and (6.8) we have neglected small symmetry
breaking effects originating from the $\epsilon^\prime$ term
in (2.12). Taking matrix elements of these equations
between nucleon states we may read off the corresponding
coupling constants at zero momentum transfer:
\begin{eqnarray}
g_{\eta^\prime NN}&=&\frac{s-t}{s}\frac{2m_N}{\sqrt3 F_\pi}H(0)
\nonumber \\
g_{GNN}&=&\frac{t}{t-s}\frac{2\sqrt3 F_\pi g_{\eta^\prime NN}}{\kappa}.
\end{eqnarray}
Next we may use the coupling constants to reexpress the axial
singlet current form factor:
\begin{eqnarray}
H(0)=\frac{\sqrt3 F_\pi}{2m_N}(g_{\eta^\prime NN}
-\frac{\kappa}{2\sqrt3  F_\pi}g_{GNN})=
``{\rm matter}"+``{\rm glue}"
\end{eqnarray}
The modifications
(6.3) and (6.6) of course enter the evaluation of the non-strange
moment of inertia $\alpha^2$, however we find, even for a large deviation
of $s-t$ from unity, no significant change of $\alpha^2$. This is
due to the fact that these modifications only affect the $\eta$ and
glueball fields  but  leave the vector meson excitations
$\xi_1,\xi_2$ and $\Phi$ essentially unaltered. Since the $\eta$ fields
do not explicitly appear in the definition of $\tilde J_\mu^5$ the
corresponding matrix element numerically turns out to be independent of
 $t$;  we find for the best fit parameters of section 5
(see table 5.3):
\begin{eqnarray}
H(0)=0.29s.
\end{eqnarray}
However, we do find a significant dependence of the predicted value for
the neutron proton mass difference, $\Delta$ on $s$ and $t$ since there the
$\eta$ fields enter crucially. In table 6.1 we display our numerical
results for $\Delta$ using the ``baryon" best fit parameters.
We note that for $|t|\le1.5$
the mass difference depends linearly on $t$ while for larger values
of $|t|$,  non-linear effects which enter via the equations of motion
are significant.
\begin{table}
\centerline{
\begin{tabular}{|c|c|c|c|c|c|}
\hline
$s \setminus t$ & 0.0 & -0.5 & -1.0 & -1.5 & -2.0  \\
\hline
0.8               &1.70 &1.84  &1.98  &2.13  &2.25     \\
\hline
0.9               &1.74 &1.87  &2.02  &2.16  &2.27     \\
\hline
1.0               &1.77 &1.91  &2.05  &2.20  &2.29     \\
\hline
1.1               &1.80 &1.95  &2.09  &2.24  &2.32     \\
\hline
1.2               &1.85 &1.98  &2.13  &2.27  &2.35     \\
\hline
\end{tabular}}
\caption{The neutron proton mass difference $\Delta$ as a function of the
parameters $s$ and $t$.}
\end{table}

To proceed, we consider the precise value of $H(0)$ as the
parameter which determines the ``fudge factor" $s$ via
eqn (6.11). Then fixing $t$ from the neutron proton mass difference
allows us to separate the matter and gluon contribution to $H(0)$ in
eqn (6.10). We present our results for various experimentally allowed
values of $H(0)$ in table 6.2 for the parameter set used to fit
the baryon mass difference as well as for the central values
of $\tilde h$ and $\tilde g_{VV\phi}$,  (5.18) together with (3.15).
\begin{table}
\centerline{
\begin{tabular}{|c|c|c|r|}
\hline
$H(0)$ & $t$ &``matter" & ``glue" \\
\hline
\hline
&\omit&\omit  Baryon Best fit & \\
\hline
0.0    &$-1.86\pm0.90$ &$0.54\pm0.26$ &$-0.54\mp0.26$ \\
\hline
0.1    &$-1.64\pm1.02$ &$0.57\pm0.30$ &$-0.47\mp0.30$ \\
\hline
0.2    &$-1.31\pm1.04$ &$0.58\pm0.30$ &$-0.38\mp0.30$ \\
\hline
0.3    &$-0.96\pm1.00$ &$0.58\pm0.29$ &$-0.28\mp0.29$ \\
\hline
\hline
&\omit&\omit  Set (3.15,5.18) & \\
\hline
0.0    &$-1.09\pm0.94$ &$0.32\pm0.26$ &$-0.32\mp0.27$ \\
\hline
0.1    &$-0.74\pm0.93$ &$0.31\pm0.30$ &$-0.21\mp0.27$ \\
\hline
0.2    &$-0.38\pm0.96$ &$0.31\pm0.30$ &$-0.11\mp0.28$ \\
\hline
0.3    &$-0.00\pm0.97$ &$0.30\pm0.29$ &$0.00\mp0.28$ \\
\hline
\end{tabular}}
\caption{The ``matter" and ``glue" contribution to the axial singlet
matrix element of the proton for two sets of parameters. For the best
fit parameters see table 5.3.}
\end{table}
We think that the results of the ``baryon" best fit should be
more reliable than the results  for the set which gives a best
fit to the meson properties. This is because for both the spin
and isospin mass splittings, a fine-tuning of the non-strange moment
of inertia $\alpha^2$ turns out to be crucial (see section 5). In any
event, the difference between the two sets provides a measure of the
``systematic" uncertainties in our calculation. Taking into
account that for the set (3.15,5.18) the absolute values for the
gluon contribution to the axial singlet matrix element are lower
than for the baryon best fit we might consider the results for the
latter set as an upper bound. We should furthermore remark that it would be
surprising if $s$ were to deviate too much from unity.  Thus, in each
case the range for $H(0)$ between 0.2 and 0.3 would seem most reasonable.

To sum up, we may conclude from table 6.2 and the discussion above that while
there is a tendency for the ``glue" part to cancel the ``matter" part,
both are most likely to be on the small side compared to unity. This agrees
with the results of [3], though it must be remarked that the present, more
detailed, calculation allows larger uncertainties in the ``glue" and
``matter" pieces.  The reason for this is that, as pointed out in the
previous section, the dependence of $\Delta$ on the $\eta$ fields is
lessened due to the inclusion of the $\gamma^\prime$ term in (2.10).  This
translates into somewhat larger values of $(s-t)$ than were previously
estimated [3] as well as large uncertainties in $t$.  Since the present
approach ``lives off" the deviation of the calculated $\Delta$ from its
``experimental" value, its accuracy could be improved if a more accurate
value of the photon exchange contribution, $\Delta_{EM}$ were available.

\vskip.3cm

\stepcounter{chapter}
\setcounter{equation}{0}
\setcounter{equation}{0}
\setcounter{footnote}{7}
\leftline{\large\it 7. Remarks on the mesonic Lagrangian}
\vskip.5cm

The effective Lagrangian discussed in sections 2$-$4 has a number of
connections  with physical quantities outside the Skyrme model approach.
We would like to make some, partly speculative, remarks on this here.
\vskip.3cm
{\it{ {\large a.} Light Quark Masses }}

Their values are of interest, for example, in constructing models of the
quark mixing (Kobayashi-Maskawa) matrix of weak interactions. Conventionally,
[13] the absolute values of the (current) quark masses are considered to be
``running" masses evaluated at the scale of 1 GeV. Estimates of the kind
we have made here yield just their ratios. Evaluation of ``QCD sum rules"
yields [13] an absolute value for $m_s(1\,\rm{GeV})$ in the 125-225 MeV range.
Combining this with the ``best fit" in (3.15) yields the sets (all in MeV)
given in table 7.1.

\begin{table}
\centerline{
\begin{tabular}{|c|c|c|}
\hline
$m_u$ & $m_d$ & $m_s$ \\
\hline
 2.3  &  5.6  &  125  \\
 3.2  & 7.9   &  175  \\
 4.2  & 10.2  &  225  \\
\hline
\end{tabular}}
\caption{Light quark masses.}
\end{table}

Note that ratios like $\sqrt{m_d/m_s}$ = 0.21 are independent of the absolute
value of\hfil\break
$m_s$(1\,GeV).

\vskip.5cm
{\it{ {\large b.} Confirmation of $R$ }}

It is interesting to try to confirm the value $R$=36 in (3.15) from
consideration of other particle multiplets.  The most precisely known masses
are, of course, those of the ordinary octet baryons.  First order perturbation
theory (which yields the fairly well satisfied Gell-Mann Okubo mass formula)
gives the prediction:
\begin{eqnarray}
R = {{\Xi-\Sigma}\over{n-p}} = {{\Sigma-N}\over{\Xi^- - \Xi^{\circ}}} \>,
\end{eqnarray}
wherein each particle symbol stands for its mass, which in turn is assumed
to have been ``corrected" by subtraction of the photon-exchange contribution.
Numerically this reads $R$ = 60.7 $\pm$ 9.1 = 46.2 $\pm$ 5.9 .  The
large deviation between the two central values suggests a need to go
beyond first order perturbation theory.  If one expands around
$m_u=m_d=m_s=0$ the procedure [13] is not straightforward since one encounters
non-analytic terms like $m^{3/2}$ and $m\ln m$. A possible way out is to
expand around the point $m_u=m_d=m_s={1\over3}(m_u+m_d+m_s)$ rather than
zero.  It was recently shown [35] that this method leads to a second order
prediction for $R$:
\begin{eqnarray}\label{eq:NewR}
R= {{3\Lambda + \Sigma -2N -2\Xi}\over{2\sqrt{3}m_T + (n-p)+
(\Xi^{\circ}-\Xi^-)}} \>,
\end{eqnarray}
where $m_T$ is the ``corrected" $\Lambda-\Sigma^{\circ}$ transition mass.
In principle $m_T$ could be determined by a precision measurement of the
difference between the $pK^-\rightarrow\Lambda\eta$ and
$n{\overline {K}}^{\circ} \rightarrow \Lambda\eta $ scattering amplitudes, for
example.  In the absence of such information we can use (\ref{eq:NewR}) to
construct a lower bound for $R$ if we assume that quantities at second
order should not deviate too drastically from their first order values.  This
yields [35] roughly
\begin{eqnarray}\label{Rlower}
R > 38 \pm 10 \>,
\end{eqnarray}
which is evidently consistent with $R$=36.

\vskip.5cm
{\it{ {\large c.} Kaplan-Manohar ambiguity }}

If ${\cal M}(x)$ is a $3\times 3$ matrix field transforming as $(3,3^*)$ under
$SU(3)_L \times SU(3)_R$, it is easy to see that $[{\cal M}(x)]^{\dagger -1}
\det {\cal M}^{\dagger}(x)$ transforms in the same way. This suggests
investigating [26] the substitution
\begin{eqnarray}\label{eq:reparam}
{\cal M} = {\cal M}^\prime + b{\cal M}^{\prime -1}\det {\cal M}^\prime \>,
\end{eqnarray}
where $b$ is a real parameter and ${\cal M}={\cal M}^\dagger$ is the diagonal
matrix proportional to quark masses, given in (2.7).  Using the characteristic
equation for a $3\times 3$ matrix it can be seen that the effect of the
substitution (\ref{eq:reparam}) on the non-derivative symmetry breakers of the
second order CPT program [12] is simply
\begin{eqnarray}\label{eq:2ndorder}
&&\delta^\prime Tr \left[{\cal M}(\tilde U + \tilde U^\dagger)\right]
 + \lambda^{'2} Tr({\cal M}\tilde U {\cal M}\tilde U + {\cal M}\tilde U^\dagger
{\cal M}\tilde U^\dagger) \phantom{aaa} \nonumber \\
&&+c_6 [Tr({\cal M}(\tilde U +\tilde U^\dagger))]^2 + c_7 [Tr({\cal M}
(\tilde U - \tilde U^\dagger))]^2  \nonumber \\
&&= \delta^\prime Tr \left[{\cal M^\prime}(\tilde U + \tilde U^\dagger)\right]
 + (\lambda^{'2}-{b\delta^\prime\over 2})
Tr({\cal M^\prime}\tilde U {\cal M^\prime}\tilde U
+ {\cal M^\prime}\tilde U^\dagger {\cal M^\prime}\tilde U^\dagger) \nonumber \\
&&+(c_6+{b\delta^\prime\over 4})
[Tr({\cal M^\prime}(\tilde U +\tilde U^\dagger))]^2 +
(c_7+{b\delta^\prime\over 4}) [Tr({\cal M^\prime}(\tilde U -
\tilde U^\dagger))]^2  + {\rm {higher \> order}}.
\end{eqnarray}
Here we have chosen to use the octet chiral field $\tilde U$, with
$\det \tilde U = 1$, for simplicity.  By ``higher order" we mean the
terms proportional to $b$ which are generated from the $\lambda^{\prime 2}$,
$c_6$ and $c_7$ terms. These are third order in the CPT program and
are to be thrown away when one is working at second order.  The significance
of (\ref{eq:2ndorder}) is that, assuming the higher order terms to be
negligible,
the identical physics is obtained using either the parameter set
$\{ m_u,m_d,m_s,\lambda^{\prime 2},c_6, c_7 \}$ or the set
$\{ m_u^\prime,m_d^\prime,m_s^\prime,(\lambda^{\prime 2})^\prime,
c_6^\prime, c_7^\prime \}$; these are related by
\begin{eqnarray}\label{eq:newset}
m_u &=& m_u^\prime + b\, m_d^\prime m_s^\prime \>, \nonumber \\
m_d &=& m_d^\prime + b\, m_u^\prime m_s^\prime \>, \nonumber \\
m_s &=& m_s^\prime + b\, m_u^\prime m_d^\prime \>, \nonumber \\
\lambda^{\prime 2} &=&  (\lambda^{\prime 2})^\prime - {b\delta^\prime \over 2}
\>, \nonumber \\
c_6 &=& c_6^\prime +  {b\delta^\prime \over 4} \>, \nonumber \\
c_7 &=& c_7^\prime +  {b\delta^\prime \over 4} \>,
\end{eqnarray}
for any $b$.

Evidently there is a one parameter family of ambiguities.  A lot of attention
has been focussed on the possibility of obtaining $m_u=0$ in some ``frame"
since that might provide a non-axionic solution of the ``$\theta$-problem".
Assuming that a solution for the parameters is specified for the primed frame
in (\ref{eq:newset}), we can transform to $m_u=0$ with the choice
$b=-m_u^\prime /(m_d^\prime m_s^\prime)$. The corresponding transformation
for $R$ in (2.9) is then
\begin{eqnarray}
R \approx R^\prime /(1+ m_u^\prime / m_d^\prime).
\end{eqnarray}
With the best fit choice (3.15) for the primed frame we find that
$R \approx 26$ is needed for $m_u=0$; according to  (\ref{Rlower}) this
would perhaps be unlikely.

The present model is, of course, different from the usual CPT program
since, among other things, we are including the vectors.  In any event,
as discussed in section 2c, the assumption that OZI rule violation is dominant
for the pseudoscalar singlet channel leads to a type 7  but not type 6
symmetry breaker.  This would uniquely fix the ``frame" for the present
purpose.

\vskip.5cm
{\it{ {\large d.} Extension of the meson Lagrangian }}

It is naturally of interest to consider how the present mesonic Lagrangian
can be modified to yield a systematic continuation of the CPT program to
higer energies.  The CPT program [12] itself is most reliable for energies
up to several hundred MeV, which restricts us to using just the pion fields.
Each successive order of approximation adds terms with two more derivatives
or one more power of the mass matrix (formally introduced as counterterms
to loops computed at the previous order). In practice, going beyond second
order may be difficult.  Perhaps surprisingly, the three flavor continuation,
including the $K$'s and the $\eta$, seems to work reasonably well.  However,
it obviously cannot be extended in energy past the vector mesons without
including them.  Since we would like to use the meson Lagrangian to study
baryons-as-solitons in the 1 GeV region, it would seem reasonable to add just
the vectors (together, perhaps, with pseudoscalar and scalar singlet glue
fields in order to develop a direct connection with QCD) and stop there.
This provides a ``clean break" in the sense of including just the lowest
lying $s$ wave states of the quark model.
To model QCD in this way to {\it {all}}
energies would likely require us to include {\it {all}} the higher resonances,
as suggested either by their need to produce high energy Regge behavior [36]
or by their presence in the large $N_c$ approximation [37].

Once we have agreed to limit the ``quarkonium" states to the pseudoscalars
and vectors it is necessary to list the chiral invariant interaction terms.
One question is whether this will introduce double counting since it is
known [38] that ``integrating out" the vectors reproduces the coefficients
of some of the pure pseudoscalar terms.  We think that this should not
be a problem; one should include {\it {a priori}} both the vector and
pseudoscalar terms and suitably readjust the coefficients of the pseudoscalar
terms.  For the ``second order" treatment, loops computed with the
``first order" terms ( (2.4) and the $\alpha^\prime$ and $\delta^\prime$
terms in (2.10) ) should be included.  For a start, however, it seems
reasonable to neglect them (while loop diagrams are of great conceptual
interest they are often numerically negligible for the indicated scale
choice {\footnote {For example, the
``chiral logs" are not qualitatively important in (10.11) of
the first of refs. [12].}}).

Here we have included those terms which we believe to be dominant. Especially,
we have neglected OZI rule violating terms, except as discussed in section
2c. The fit to experiment is reasonable for the 2 and 3 point functions
considered,  Trying to fit more quantities will probably require fine
tuning by adding additional terms and including chiral logs.  Considering
the tremendous amount of data in the region below 1 GeV, it is clear that
the best way to proceed is along an incremental ``evolutionary" path.

\vskip.5cm
\leftline{\large\it Acknowledgements}
\vskip.3cm

We would like to thank Hans Walliser for a helpful discussion. This work
was supported in part by the U.S. Department of Energy under contract
number DE-FG-02-85ER40231 and in part by the Deutsche
Forschungsgemeinschaft (DFG) under contract Re 856/2-1.

\vskip.5cm

\appendix

\stepcounter{chapter}
\setcounter{equation}{0}
\leftline{\large\it Appendix}
\vskip.3cm

In this appendix we list the analytic expressions for the isospin
breaking parameters in the collective Lagrangian (5.9).

Substituting the classical fields into the symmetry breaking part of
the Lagrangian (2.10) yields:
\begin{eqnarray}
\Gamma_3&=&{{8\pi y}\over{\sqrt3}}\int dr  \tilde \Gamma_3 \nonumber \\
\tilde \Gamma_3&=&{{\alpha^\prime}\over{2\tilde g^2}}
\big[r^2{\rm cos}F(F^{\prime2}+\omega^2)
-2G(G+2){\rm cos}F+4(1+G-{\rm cos}F)\big] \nonumber \\
&&-2\beta^\prime{\rm cos}F(F^{\prime2}r^2+2{\rm sin}^2F)
+2\delta^\prime r^2(1-{\rm cos}F) \nonumber \\
&&-{{\gamma^\prime}\over{\tilde g^2}}{\rm cos}F(2G^{\prime2}+
{{G^2}\over{r^2}}(G+2)^2-\omega^{\prime2}r^2) \nonumber \\
&&+{{4(2+x)\lambda^{\prime2}}\over{3}}r^2(1-{\rm cos}2F)
\end{eqnarray}
and
\begin{eqnarray}
\Delta_3={{32\pi}\over{3\sqrt3}}(1-x)y\lambda^{\prime2}
\int dr r^2 (2 {\rm sin}^2F+3{\rm cos}F-3).
\end{eqnarray}

For the isospin breaking parameters which appear in those terms which
are linear in the angular velocities we find:
\begin{eqnarray}
\alpha_3&=&{{8\pi y}\over3}\int dr \tilde{\alpha_3}
\qquad \beta_3=4\pi y\int dr \tilde{\beta_3} \nonumber \\
\tilde{\alpha_3}&=&{{\alpha^\prime}\over{\tilde g^2}}
\big\{\omega r^2\big[2-2\xi_1-{\rm cos}F(2+\xi_1+\xi_2)]
-4\Phi(1+G-{\rm cos}F) \nonumber \\
&&+(\chi+\chi_8){\rm sin}F\big[{1\over2}\omega^2r^2
+{1\over2}F^{\prime2}r^2-G(G+2)\big] \nonumber \\
&&-\big[r^2{\rm cos}FF^\prime(\chi^\prime+\chi_8^\prime)+
2(\chi+\chi_8){\rm sin}F\big]\big\} \nonumber \\
&&-2\beta^\prime\big[(\chi+\chi_8){\rm sin}F
(F^{\prime2}r^2+2{\rm sin}^2F)
-2(r^2{\rm cos}FF^\prime(\chi^\prime+\chi_8^\prime)
+2{\rm sin}F(\chi+\chi_8))\big]
\nonumber \\
&&-{{\gamma^\prime}\over{\tilde g^2}}\big[(\chi+\chi_8){\rm sin}F
\big(2G^{\prime2}+{{G^2}\over{r^2}}(G+2)^2-\omega^{\prime2}r^2\big)
\nonumber \\
&&+4(\omega^\prime\xi_1^\prime r^2+2G^\prime\Phi^\prime)
+2{\rm cos}F\big(\omega^\prime(\xi_1^\prime+\xi_2^\prime)r^2
+{4\over{r^2}}\Phi G(G+2)\big)\big] \nonumber \\
&&-2\delta^\prime r^2 (\chi+\chi_8){\rm sin}F
-2\epsilon^{\prime}r^2g{\rm sin}F
-8\frac{2+x}{3}\lambda^{\prime2}r^2 (\chi+\chi_8){\rm sin}2F \\
\tilde{\beta_3}&=&{{\alpha^\prime}\over{2\tilde g^2}}
\big\{W({\rm sin}{F\over2}+{\rm sin}F)[\omega^2r^2
-2(1+G-{\rm cos}F)^2] \nonumber \\
&&+2{\rm cos}{F\over2}\big[(1+G-{\rm cos}F)(D+2W{\rm sin}{F\over2}
(1+{\rm cos}{F\over2}))
-\omega r^2(S-1+{\rm cos}{F\over2})\big]\big\} \nonumber \\
&&+{{\alpha^\prime}\over{\tilde g^2}}\big\{r^2\omega {\rm sin}^2{F\over2}
\nonumber \\
&&-W(1+G-{\rm cos}F)
\big[{\rm sin}{F\over2}(1+2{\rm cos}{F\over2}+{\rm cos}F)
+2{\rm sin}F({\rm cos}{F\over2}+{\rm cos}F)\big] \nonumber \\
&&+{\rm sin}{F\over2}\big[\frac{r^2}{2}F^\prime
(E+2{\rm sin}{F\over2}W^\prime-F^\prime W)
-{\rm sin}F((D+2W{\rm sin}{F\over2}(1+{\rm cos}{F\over2}))\big]\big\}
\nonumber \\
&&+2\big({{\alpha^\prime}\over{4\tilde g^2}}-\beta^\prime\big)
\big\{W({\rm sin}{F\over2}+{\rm sin}F)(F^{\prime2}r^2+2{\rm sin}^2F)
\nonumber \\
&&-2{\rm cos}{F\over2}(1+{\rm cos}{F\over2})
\big[r^2F^\prime W^\prime+2{\rm cos}{F\over2}{\rm sin}FW\big]\big\}
\nonumber \\
&&-4\delta^\prime r^2 {\rm sin}{F\over2}
(1+{\rm cos}{F\over2})W \nonumber \\
&&+{{\gamma^\prime}\over{\tilde g^2}}\big\{
{\rm cos}{F\over2}\big[G^\prime(2D^\prime+EG)
+{G\over{r^2}}(G+2)^2D
-2\omega^\prime r^2(S^\prime+{{\omega}\over4}E)\big] \nonumber \\
&&-W({\rm sin}{F\over2}+{\rm sin}F)
\big[2G^{\prime2}+{{G^2}\over{r^2}}(G+2)^2
-\omega^{\prime2}r^2\big]\big\} \nonumber \\
&&-4\lambda^{\prime2}r^2W\big[{\rm sin}2F+2{\rm sin}{{3F}\over2}+
{\rm sin}F+2x{\rm sin}{F\over2}(1+{\rm cos}{F\over2})\big]
\end{eqnarray}

Of course, for the parts of the Lagrangian already investigated in ref.[1]
the expressions for $\Gamma_3$ and $\alpha_3$ coincide.

\vskip.5cm
\leftline{\large\it Footnotes}
\vskip.3cm
\begin{enumerate}
\item Only 3.8 was used for
fitting in ref. [1].  The extra factor $Z_{\phi}^{-2}$ makes negligible
difference. Note that the $\lambda^{\prime2} \rightarrow 0$ limit of (3.6)
quoted, but not used for fitting, should be replaced by the present formula.

\item Input parameters: $F_{\pi p}=0.132\, {\rm {GeV}},
F_{Kp}=0.161\, {\rm {GeV}}, m_\pi=0.137\,{\rm {GeV}},m_{K^*}=0.892\,
{\rm {GeV}},m_\rho=0.768\,{\rm {GeV}},m_K=0.497\,{\rm {GeV}},
(m_{K^\circ}-m_{K^+})_{non-EM}=5.28\,{\rm {MeV}},M_{\rho\omega}=-2.65\,
{\rm {MeV}}, \Gamma(\rho \rightarrow 2\pi)=0.1491\,{\rm {GeV}},
\Gamma(K^* \rightarrow K\pi)=0.0498\,{\rm {GeV}},
\Gamma(\phi \rightarrow K{\overline K})=0.0037\,{\rm {GeV}}$.

\item In the extraction
of the parameters $c_1$ and $c_2$ of (2.5) from experiment in [22]
some $\phi$ decays are involved.  We note that
in (4.5) of this reference we should now write $\Big|\epsilon/ Z_\phi
\Big|$ instead of $|\epsilon|$.  This replacement should
also be made in (4.8) so that the old formulas for $c_1$ and $c_2$ remain
unaltered.

\item This can be seen by referring to Fig 2 of [32].

\item More details on the notation of this section may
be found in refs [17].

\item Hans Walliser has pointed out to us that the
experimental measurements for the relevant $\omega$ and $\phi$ partial
decay widths have changed from the 1986 to the 1992 ``Review of Particle
Properties". Continuing to use (4.4) [see footnote 3 above], (4.7)
and (4.9) of [22] now gives us $|\epsilon|/Z_\phi = 0.059 \pm 0.005$
and central values $\tilde {g}_{VV\phi} = 1.81, \, \tilde{h}=0.38$.
These are substantially similar to the older values.

\item Correct charge normalization requires that we only include
the contributions of the classical fields (5.3) to $\alpha_1$ and $\beta_1$.

\item For example, the
``chiral logs" are not qualitatively important in the first of refs. [12].).

\end{enumerate}
\vskip.5cm

\leftline{\large\it References}
\vskip.3cm
\begin{enumerate}

\item P. Jain, R. Johnson, N.W. Park, J. Schechter and H. Weigel, Phys. Rev.
{\bf {D40}}, 855 (1989).

\item R. Johnson, N.W. Park, J. Schechter, V. Soni and H. Weigel, Phys. Rev.
{\bf {D42}}, 2998 (1990).

\item J. Schechter, V. Soni, A. Subbaraman and H. Weigel, Phys. Rev. Lett.
{\bf {65}}, 2955 (1990).

\item E.M. Collaboration, J. Ashman et al, Phys. Lett. {\bf {B206}},
364 (1988); Nucl. Phys. {\bf {B328}}, 1 (1989).

\item Reviews are provided by R. Jaffe and A. Manohar, Nucl. Phys. {\bf
{B337}},
509 (1990); G. Altarelli and W.J. Stirling, CERN report Th 5249/88;
H.Y. Cheng, Taipei report IP-ASTP-01-91; T.P. Cheng and L.F. Li,
Carnegie Mellon report HEP 90-2.

\item  A ``brief review" from the standpoint of the present paper is given in
J. Schechter, V. Soni, A. Subbaraman and H. Weigel, Mod. Phys. Lett {\bf {A7}},
1 (1992). More recent references include K.-F. Lin, Phys. Lett. {\bf B281},
141 (1992); M.K. Banerjee and T.D. Cohen, Maryland report 91-267; K. Kobaykawa,
T. Morii, S. Tanaka and T. Yamanishi, Kobe report 91-03; M. Wakamatsu, Osaka
report; K.-Ta Chao, J.-Ru Wen and H.-Q. Zheng, CERN report TH 6288/91.

\item S. Brodsky, J. Ellis and M. Karliner, Phys. Lett. {\bf {B206}}, 309
 (1988).

\item V. Bernard, N. Kaiser and Ulf-G. Mei\ss ner, Phys. Lett. {\bf {B237}},
545 (1990).

\item R. Carlitz, J. Collins and A. Mueller, Phys. Lett. {\bf {B214}},
229 (1988);
A.V. Efremov and O.J. Teryaev, Dubna report JINR E2-88-297; G. Altarelli and
G.G. Ross, Phys. Lett. {\bf {B212}}, 391 (1988).

\item See A. Manohar, Phys. Rev. Lett. {\bf {66}}, 1663 (1991);
J. Mandula, Phys. Rev. Lett. {\bf {65}}, 1403 (1990).

\item G.M. Shore and G. Veneziano, Phys. Lett. {\bf {B244}}, 75 (1990);
G. Veneziano, Mod. Phys. Lett. {\bf {A4}}, 1605 (1989); G.M. Shore and
G. Veneziano, CERN report TH 6019/91.

\item See J. Gasser and H. Leutwyler, Nucl. Phys. {\bf {B250}}, 465 (1985).
For earlier work consult L.-F. Li and H. Pagels, Phys. Rev. Lett. {\bf {26}},
1204 (1971); P. Langacker and H. Pagels, Phys. Rev. {\bf {D8}}, 4395 (1973);
S. Weinberg, Physica {\bf {96A}}, 327 (1979).

\item See J. Gasser and H. Leutwyler, Phys. Rep. {\bf {87}}, 77 (1982).

\item S. Okubo, Phys. Lett. {\bf {5}}, 165 (1963).

\item G. Zweig, CERN reports 8182/TH 401, 8419/TH 412 (1964); J. Iizuka,
Prog. Theor. Phys. Suppl. {\bf {37-8}}, 21 (1966).

\item H. Yabu and K. Ando, Nucl. Phys. {\bf {B301}}, 601 (1988).

\item The full excitation spectrum is discussed in N.W. Park and H. Weigel,
Phys. Lett. {\bf B268}, 420 (1991); Nucl. Phys. {\bf {A541}}, 453 (1992).

\item The strange pseudoscalar excitations were discussed in
H. Weigel, J. Schechter, N.W. Park and Ulf-G. Mei\ss ner, Phys.
Rev. {\bf {D42}}, 3177 (1990).

\item The non-strange vector excitations for a different meson
Lagrangian  were given in Ulf-G. Mei\ss ner, N. Kaiser and W. Weise,
Nucl. Phys. {\bf {A466}}, 685 (1987).

\item Ulf-G. Mei\ss ner, N. Kaiser, H. Weigel and J. Schechter,
Phys. Rev. {\bf {D39}}, 1956 (1989);

\item \"O. Kayamakcalan and J. Schechter, Phys. Rev. {\bf {D31}},1109 (1985).

\item P. Jain, R. Johnson, Ulf-G. Mei\ss ner, N.W. Park and J. Schechter, Phys.
 Rev.
{\bf {D37}}, 3252 (1988).

\item \"O. Kayamakcalan, S. Rajeev and
J. Schechter, Phys. Rev. {\bf {D30}}, 594 (1984).

\item E. Witten, Nucl. Phys. {\bf {B223}}, 422 (1983).

\item A different approach given by T. Fujiwara, T. Kugo, H. Terao,
S. Uehera and
K. Yamawaki, Prog. Theor. Phys. {\bf {73}}, 926 (1985) leads to the identical
Lagrangian when their CP violating terms are deleted.

\item D. Kaplan and A. Manohar, Phys. Rev. Lett {\bf {56}}, 2004 (1986).

\item C. Rosenzweig, J. Schechter and G. Trahern, Phys. Rev. {\bf {D21}}, 3388
(1980); P. Di Vecchia and G. Veneziano, Nucl. Phys. {\bf {B171}}, 253 (1980);
E. Witten, Ann. Phys.{\bf {128}}, 1789 (1981);
P. Nath and R. Arnowitt, Phys. Rev. {\bf {D23}}, 473 (1981).

\item P. Di Vecchia, F. Nicodemi, R. Pettorino and G. Veneziano, Nucl. Phys.
{\bf {B181}}, 318 (1981).

\item J. Donoghue and D. Wyler, Phys. Rev. {\bf {D45}}, 892 (1992).

\item See table X of N. Deshpande, D. Dicus, K. Johnson and V. Teplitz,
Phys. Rev. {\bf {D15}}, 1885 (1977).

\item Review of Particle Properties, Particle Data Group, Phys. Rev.
{\bf D45}, S1 (1992).

\item V. Mirelli and J. Schechter, Phys. Rev. {\bf {D15}}, 1361 (1977).

\item C. Callan and  I. Klebanov, Nucl. Phys. {\bf B262},
365 (1985); C. Callan, K. Hornbostel and I. Klebanov, Phys. Lett.
{\bf B202}, 296 (1988);
I. Klebanov in $Hadrons$ $and$ $Hadronic$ $Matter$, page 223, proceedings of
the NATO Advanced Study Institute, Cargese, 1989, edited by D. Vautherin,
J. Negele and F. Lenz (Plenum Press 1989).

\item See ref. [16] and appendix A of N.W. Park, J. Schechter and
H. Weigel, Phys. Rev. {\bf D43}, 869 (1991).

\item J. Schechter and A. Subbaraman, Int. Jour. Mod. Phys. A, to be published.

\item See, for example, L. Van Hove, Phys. Letts. {\bf 24B}, 183 (1967).

\item See, for example, G. 't Hooft, Nucl. Phys. {\bf B72}, 461 (1974);
E. Witten, Nucl. Phys. {\bf B160}, 57 (1979).

\item See, for example, J. Donoghue, C. Ramirez and G. Valencia, Phys. Rev.
{\bf D39}, 1947 (1989); G. Ecker, J.Gasser, A. Pich and E. de Rafael,
Nucl. Phys. {\bf B221}, 311 (1989).

\end{enumerate}

\end{document}